\documentclass[aps,prd,twocolumn,nofootinbib,superscriptaddress]{revtex4-2}
\usepackage[utf8]{inputenc}
\usepackage{orcidlink}
\usepackage{color}
\usepackage{graphicx}
\usepackage{amsmath}
\usepackage{amssymb}
\usepackage{bm}
\usepackage{acronym}
\usepackage{ifthen}
\usepackage{blindtext}
\usepackage[normalem]{ulem}
\usepackage{hyperref}
\usepackage{etoolbox}
\usepackage{fancyhdr}
\usepackage{xspace}
\usepackage{textcomp}
\usepackage{multirow}
\usepackage[caption=false]{subfig}
\usepackage{tabularx}

\newacro{S/N}{signal-to-noise ratio}

\newacro{PN}{post-Newtonian}

\newacro{O3}{the third observing run}

\newacro{O2}{the second observing run}

\newacro{CW}{\emph{Continuous Wave}}

\newcommand{\bea}{\begin{eqnarray}}
\newcommand{\eea}{\end{eqnarray}}
\newcommand{\be}{\begin{equation}}
\newcommand{\ee}{\end{equation}}

\newtoggle{fullauthorlist}
\toggletrue{fullauthorlist}
\newtoggle{endauthorlist}
\toggletrue{endauthorlist}

\begin{document}
\title{A semi-coherent generalization of the 5-vector method to search for continuous gravitational waves}
\author{Sabrina D'Antonio}
\affiliation{INFN, Sezione di Roma Tor Vergata, I-00133 Roma, Italy}
\author{Cristiano Palomba}
\affiliation{INFN, Sezione di Roma, I-00185 Roma, Italy}
\author{Pia Astone}
\affiliation{INFN, Sezione di Roma, I-00185 Roma, Italy}
\author{Simone Dall'Osso}
\affiliation{INFN, Sezione di Roma, I-00185 Roma, Italy}
\author{Stefano Dal Pra}
\affiliation{INFN-CNAF, I-40127 Bologna, Italy}
\author{Sergio Frasca}
\affiliation{Universit\`a di Roma ``La Sapienza'', I-00185 Roma, Italy}
\author{Paola Leaci}
\affiliation{Universit\`a di Roma ``La Sapienza'', I-00185 Roma, Italy}
\affiliation{INFN, Sezione di Roma, I-00185 Roma, Italy}
\author{Federico Muciaccia}
\affiliation{INFN, Sezione di Roma, I-00185 Roma, Italy}
\author{Ornella J Piccinni}
\affiliation{Institut de F\'{\i}sica d'Altes Energies (IFAE), E-08193 Barcelona, Spain }
\author{Lorenzo Pierini}
\affiliation{Universit\`a di Roma ``La Sapienza'', I-00185 Roma, Italy}
\affiliation{INFN, Sezione di Roma, I-00185 Roma, Italy}
\author{Marco Serra}
\affiliation{INFN, Sezione di Roma, I-00185 Roma, Italy}

\begin{abstract}
The emission of continuous gravitational waves (CWs), with duration much longer than the typical data taking runs, is expected from several sources, notably spinning neutron stars, asymmetric with respect to their rotation axis and more exotic sources, like ultra-light scalar boson clouds formed around Kerr black holes and sub-solar mass primordial binary black holes. Unless the signal time evolution is well predicted and its relevant parameters accurately known, the search for CWs is typically based on semi-coherent methods, where the full data set is divided in shorter chunks of given duration, which are properly processed, and then incoherently combined. 


In this paper we present a semi-coherent method, in which the so-called \textit{5-vector} statistics is computed for the various data segments and then summed after the removal of the Earth Doppler modulation and signal intrinsic spin-down. The method can work with segment duration of several days, thanks to a double stage procedure in which an initial rough correction of the  Doppler and spin-down is followed by a refined step in which the residual variations are removed. 

This method can be efficiently applied for directed searches, where the source position is known to a good level of accuracy, and in the candidate follow-up stage of wide-parameter space searches. 
\end{abstract}

\maketitle
\section{Introduction}


Gravitational wave astronomy started in 2015 with the detection of gravitational waves (GWs) emitted in the last stages of the coalescence of a black hole binary system \cite{2016PhRvL.116f1102A}. To date, a total of 90 binary events have been observed by the LIGO \cite{2015CQGra..32g4001L} and Virgo detectors \cite{2015CQGra..32b4001A}, all due to the coalescence of compact binaries made, in the vast majority of the cases, by a pair of black holes and, in very few cases, by a pair of neutron stars or by a black hole and a neutron star \cite{2021arXiv211103606T} . 

However, many more kinds of sources are expected to exist, see e.g. \cite{2015PASA...32...34L} \cite{2018ASSL..457..673G} \cite{PhysRevD.103.083020} for recent reviews. In particular, we are interested in CW sources, which emit signals with duration longer than one day, for which the modulations due to the Earth rotation play a critical role. This kind of emission characterizes, for instance, spinning neutron stars asymmetric with respect to the axis of rotation. Asymmetric spinning neutron stars, isolated or in a binary system, are considered the prototypical source of CWs. Their detection will be a fundamental milestone in GW physics because they can be observed for very long times, becoming true laboratories for fundamental physics and astrophysics.
Recently, more exotic sources of CWs have been also proposed which, if detected, would shed light on several important aspects of fundamental physics and cosmology, including dark matter. One example is represented by ultra-light boson clouds that may form around Kerr black holes, as a consequence of a \textit{superradiance} process \cite{Arvanitaki:2010sy} \cite{Yuan:2021ebu}. Once formed, the cloud will dissipate through the emission of a CW signal, with a secular spin-up in frequency. Another interesting example are binary systems made of sub-solar mass primordial black holes \cite{raidal2019formation} \cite{clesse2018seven}. Such systems, for values of the chirp mass smaller than $\sim 10^{-3}$ solar masses, are characterized by a very long coalescence time, and thus emit a nearly periodic signal with a slowly increasing frequency.

The search for CWs can be based on optimal fully coherent methods (using matched filtering), only when the source sky position, frequency and frequency evolution are accurately known, see e.g. \cite{Jaranowski:1998qm} \cite{2010CQGra..27s4016A}. Otherwise, regardless of the source, the search is computationally very heavy and relies on semi-coherent approaches that strongly reduce the required computing power - with respect to matched filtering - at the price of a sensitivity loss
(see e.g. \cite{PhysRevD.57.2101,PhysRevD.61.082001,Krishnan:2004sv, Suvorova:2016rdc,2017PhRvD..95l2001L, Dergachev:2010tm, AsDaFrPa:2014_PRD,Aasi:2014, 2021PhRvD.103f3030D,2021ApJ...909...79S,PhysRevD.72042004,PhysRevD.94064061,PhysRevD.92.082003,PhysRevD.94.122002}). 
Several such methods have been applied to LIGO-Virgo data from various runs, see e.g. \cite{2019PhRvD..99l2002A, 2021PhRvD.103f3019D, 2021PhRvD.103f4017A, 2021ApJ...909...79S,2022PhRvD.106j2008A, 2022ApJ...932..133A,2021ApJ...909...79S} for recent results concerning all-sky searches, and \cite{tenorio2021search,piccinni2022status,riles2023searches} for general reviews on CW search methods. Generally speaking, in a semi-coherent method the full data set is divided in several shorter chunks that are independently processed, and then re-combined incoherently, i.e. without taking into account the signal phase.
The various approaches differ under different aspects: the segment length, the way in which each data segment is processed, the statistics used to measure the significance of the results, the way in which noise artefacts are dealt with, and in several implementation details. In any case, the goal is an analysis method which is as sensitive, robust and computationally cheap as possible. 

In this paper we introduce a new semi-coherent method of analysis that exploits the sidereal amplitude modulation of CW signals, induced by the  time-varying response 
of the detector. The method is built on the so-called \textit{5-vector} statistics \cite{2010CQGra..27s4016A}, largely used in targeted searches for known pulsars, which is here adapted to a semi-coherent scheme. This new pipeline allows to make sensitive and computationally cheap searches, with coherence time of several sidereal days, toward specific sky directions. As such, it can be used, for instance, to make directed searches toward globular clusters or the galactic center and for the follow-up of outliers found in wide-parameter space searches (like all-sky searches). \\ 

The paper is organized as follows. In Sec. \ref{sec:scheme} a brief introductory description of the method is given. In Sec. \ref{sec:tfft} the computation of the coarse frequency grid is described. In Sec. \ref{sec:fivevec} the \textit{5-vector} statistics is briefly reminded, and its use in the context of a semi-coherent method discussed. Sec. \ref{sec:semicoh} is devoted to outline the removal of the residual Doppler effect, and the computation of the semi-coherent statistics. Sec. \ref{sec:spindown} extends the algorithm to the presence of a source intrinsic spin-down. Sec. \ref{sec:chara} describes the experimental procedure used to estimate the method sensitivity, including a comparison with the theoretical computation, discusses some implementation details of the analysis procedure and, finally, briefly comments on the computational cost of the algorithm. In Sec. \ref{sec:applica} the validation tests done with hardware and software simulated signals are discussed. Finally, Sec. \ref{sec:concl} contains the conclusions. Details on the theoretical sensitivity computation are given in Appendix \ref{theosens}. 

\section{Overview of the method}
\label{sec:scheme}
\begin{figure}
    \centering
    \includegraphics[width=0.8\columnwidth]{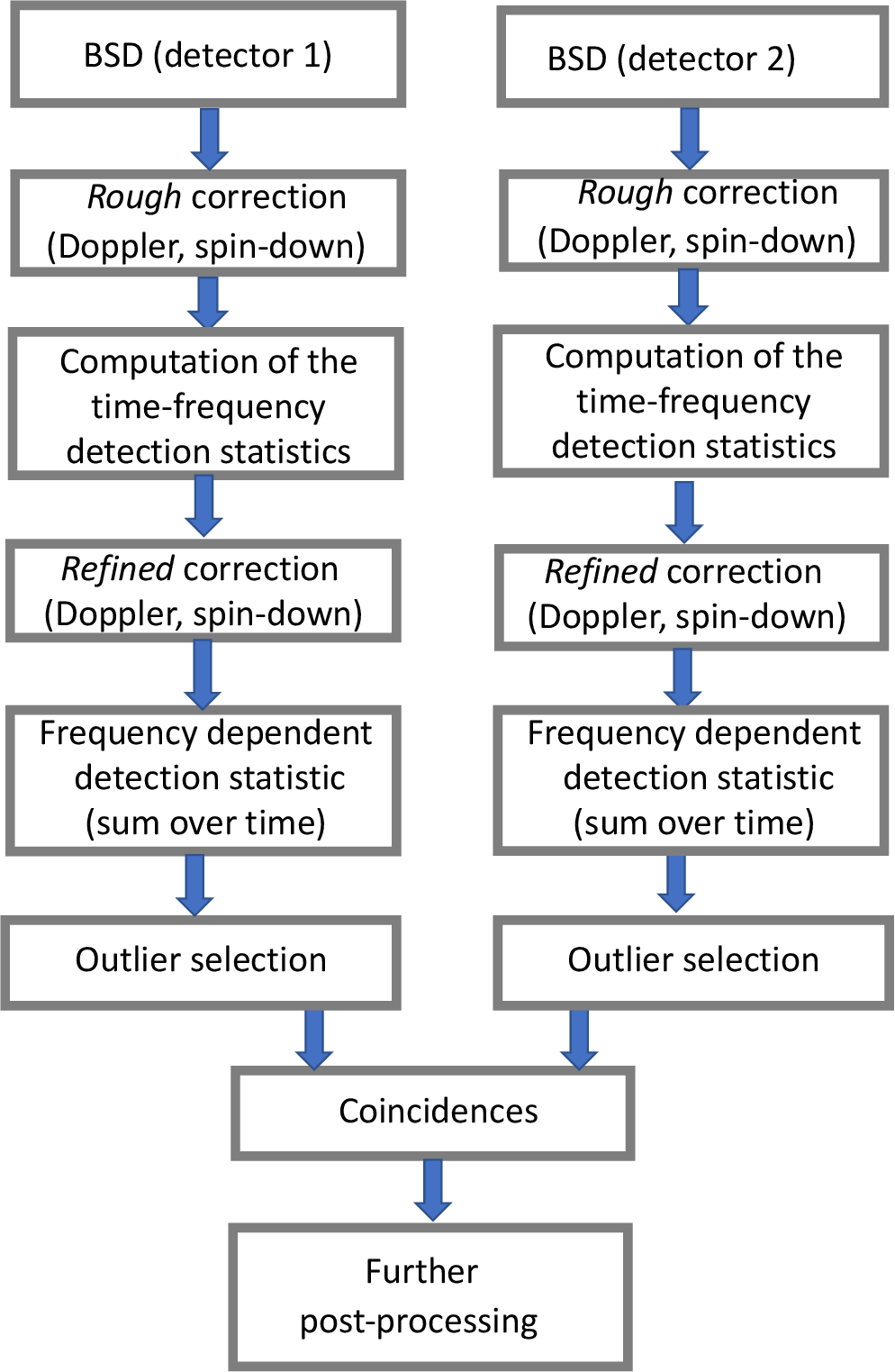}
    \caption{Scheme of the analysis method, assuming the analysis is done toward a single sky direction and that data from two detectors are used. See text for more details.}
    \label{fig:analysis_scheme}
\end{figure}

In this section we give an overview of the analysis method, leaving details to following sections. A scheme of the method is shown in Fig. \ref{fig:analysis_scheme} where, as an example, data from two detectors are considered and a fixed sky direction is assumed.
The starting point is represented by BSD (Band Sampled Data) files containing detector calibrated data, each file covering 1 month and a band of 10 Hz, and cleaned from short duration disturbances \cite{piccinibsd}. 
For a given target of the search to be performed, a \textit{coarse} grid on the parameter space, consisting of frequency, spin-down and possibly sky position, is built. 
The range covered by each parameter depends on the specific target. For instance, for the follow-up of an all-sky outlier it will consist of a small interval around the outlier parameters, determined by the uncertainty associated to each of them. In the case of a directed search toward, e.g., a globular cluster, we will likely employ a large range of  values for frequency and spin-down, which are typically unknown, and just one, or a few, sky position(s) in order to cover the globular cluster extension. 

For each dataset the data time series is subject to a heterodyne correction \cite{piccinibsd} of the Doppler modulation (Sec. \ref{sec:tfft}) and intrinsic source spin-down (Sec. \ref{sec:spindown}), done over the \textit{coarse} grid points. This allows for a partial substraction of those frequency variations. The grid is built in such a way that, over data segments of duration $T_{\rm FFT}$, any residual frequency variation is confined within one frequency bin of width $\delta f=1/T_{\rm FFT}$. The choice of $T_{\rm FFT}$, and then the corresponding number of grid points, is a matter of compromise between sensitivity (longer $T_{\rm FFT}$) and computational cost (higher number of grid points). When a large frequency range is considered, for practical reasons, this is split in several smaller bands, say 1 Hz wide, and the analysis steps described in the following are repeated for each of them. 
For each time segment of duration $T_{\rm FFT}$, the \textit{5-vector} statistics is computed (Sec. \ref{sec:fivevec}) for each frequency bin (within a 1 Hz subband) and a time-frequency map of the statistics values is built. The residual variation in frequency and spin-down is then removed by building a \textit{refined} grid and applying the needed corrections by properly shifting the frequency bins in the time-frequency plane (Secs. \ref{sec:semicoh} and \ref{sec:spindown}). At this point, a total statistics is built by summing the statistics computed over each segment. Finally, a fixed number of the most significant outliers are selected in each 1 Hz subband, over the whole sky and spin-down ranges \cite{2022PhRvD.106j2008A}. The outliers are subject to subsequent analyses steps, which have been commonly used in previous searches \cite{2022PhRvD.106j2008A} and for this reason are not discussed in detail in this paper. In brief, coincidences among outliers found in different datasets (see penultimate box in Fig. \ref{fig:analysis_scheme}) consist in initially taking only those which \textit{distance} in the parameter space is below some pre-defined threshold, ranking them on the base of a combination of distance and Critical Ratio, and keeping a given number of the most significant among these, see e.g. \cite{2014PhRvD..89f2008A} for a detailed discussion. These surviving outliers are subject to additional post-processing \cite{2022PhRvD.106j2008A} 
\cite {PhysRevD.104.084012} \cite{PhysRevD.104.084012} to discard those which are not compatible with an astrophysical signal. Further iterations of the semi-coherent procedure are possibly applied to deeply follow any remaining candidate.  
This method represents the evolution and improvement of a previous simpler approach used in \cite{Abbott_2021} \cite{PhysRevD.106.042003}.\\ 

\section{\textit{Coarse} frequency grid}
\label{sec:tfft}
In this section we describe how the \textit{coarse} frequency grid is built, deferring the discussion on spin-down correction to Sec. \ref{sec:spindown} and assuming a fixed sky position. At each point of this grid an hetherodyne correction is applied in order to partially remove the Doppler effect.

A CW signal before reaching the detector can be represented, in complex notation, as $h(t)=h_0e^{i \phi(t)}$.
At the detector, the signal is characterized by an amplitude modulation, not relevant here and discussed in Sec. \ref{sec:fivevec}, and a frequency modulation. The corresponding phase evolution for a signal with intrinsic frequency $f_0$, and assuming for simplicity zero spin-down (an assumption which will be relaxed later), can be expressed as
\be
\label{Eq0}
\phi(t)=\phi_\mathrm{0}+\omega_\mathrm{0}\left(t+\frac{\vec{r}_{\rm t} \cdot \hat{n}}{c}\right ),
\ee
where $\omega_0=2 \pi f_0$, $\vec{n}$ is the unit vector identifying the direction to the source, $\vec{r}_{\rm t}$ is the time-dependent detector position in the reference frame of the solar system barycenter (SSB). The Romer delay $\frac{\vec{r}_{\rm t} \cdot \hat{n}}{c}$ is responsible for the Doppler effect due to the motion of the detector. 

\begin{figure}
    \centering
    \includegraphics[width=\columnwidth]{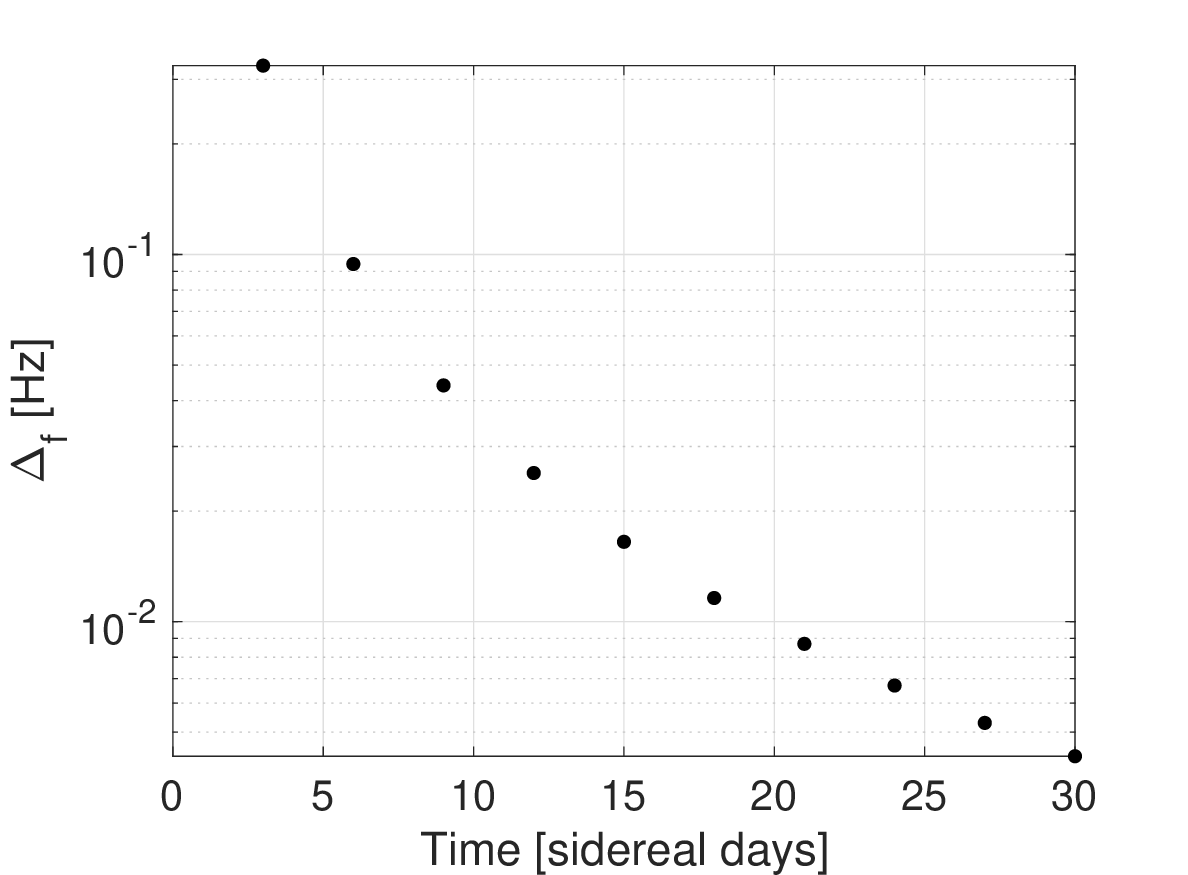}
    \caption{Frequency coarse step, $\Delta_f$, as a function of data segment duration $T_{\rm FFT}$, for a source direction $(\lambda=193.3162^o,~\beta=-30.9956^o)$.}
    \label{fig:Step_vs_tfft}
\end{figure}

If we perform an etherodyne correction over the whole dataset to compensate the Doppler effect, using the right sky position and a wrong angular frequency $\omega_\mathrm{0}^{'}$, i.e. we multiply the data by a factor $e^{-i\omega^{'}_\mathrm{0}\frac{\vec{r}_{\rm t} \cdot \hat{n}}{c}}$,
the resulting signal phase is
\begin{equation}
\label{Eq1}
\phi_\mathrm{corr}(t)=\omega_\mathrm{0}t+ \left(\omega_\mathrm{0}-\omega'_\mathrm{0}\right)\frac{\vec{r}_{\rm t} \cdot \hat{n}}{c}
\end{equation}
Due to the wrong correction, the resulting signal frequency is affected by a residual Doppler modulation:
\begin{equation}
\label{eq: f_variation}
f(t)=\frac{1}{2\pi}\frac{d\phi_\mathrm{corr}}{dt}=f_\mathrm{0}+(f_0-f'_\mathrm{0})\frac{\vec{v}_{\rm t} \cdot \hat{n}}{c},
\end{equation}
where $\vec{v}_{\rm t}$ is the detector velocity in the SSB reference frame and $f'_\mathrm{0} = \omega'_\mathrm{0}/2 \pi $. 
At two different times $t_\mathrm{1}$ and $t_\mathrm{2}$, the wrongly corrected and the true signal frequencies differ by
\begin{equation}
f(t_\mathrm{1})-f_\mathrm{0}=(f_\mathrm{0}-f'_\mathrm{0})\frac{\vec{v}_{\rm t_1} \cdot \hat{n}}{c}
\end{equation}
\begin{equation}
f(t_\mathrm{2})-f_\mathrm{0}=(f_\mathrm{0}-f'_\mathrm{0})\frac{\vec{v}_{\rm t_2} \cdot \hat{n}}{c}
\end{equation}
It follows that
\begin{equation}
\label{eq: f_variation12}
f(t_\mathrm{1})-f(t_\mathrm{2})=(f_\mathrm{0}-f'_\mathrm{0})\left(\frac{\vec{v}_{\rm t_2} \cdot \hat{n}}{c}-\frac{\vec{v}_{\rm t_1} \cdot \hat{n}}{c}\right) 
\end{equation}
We use this equation to define which the maximum time interval $T_{\rm FFT}=t_2-t_1$ such that the full signal power is confined within a single frequency bin $\delta f =1/(t_2-t_1)$. Specifically, the frequency variation given by Eq. \ref{eq: f_variation12} over the time interval $T_{\rm FFT}$ must meet the condition
\begin{equation}
\label{eq:condition}
\left|(f_\mathrm{0}-f'_\mathrm{0})\left(\frac{\vec{v}_{\rm t_2} \cdot
\hat{n}}{c}-\frac{\vec{v}_{\rm t_1} \cdot \hat{n}}{c}\right)\right|_\mathrm{max}\le\frac{1}{T_{\rm FFT}},
\end{equation}
where the maximum of the expression in parentheses is taken across the whole observing time. For a fixed $T_{\rm FFT}$, and for a given detector and sky position, Eq. \ref{eq:condition} provides the maximum allowed coarse frequency step $\Delta_{\rm{f}}=f_\mathrm{0}-f'_\mathrm{0}$ when looking for a CW source emitting at an unknown frequency $f_\mathrm{0}$.  
When searching over a frequency band of width $B_0$ and starting point $f_\mathrm{start}$, the grid frequencies $f'_\mathrm{0}=f_\mathrm{start} + K\cdot \Delta_{\rm{f}},~ K=1,...M$, where $M=\mathrm{round}(B_0/\Delta_{\rm{f}})$, will thus differ from $f_\mathrm{0}$ by at most $\Delta_{\rm{f}}$. It is important to note that here we are not taking into account the signal sidereal modulation, which determines an additional spread of the signal power, and that will be considered in Sec. \ref{sec:fivevec}. Through Eq.\ref{eq:condition} we can then set the values for the pair ($T_{\rm FFT},~ \Delta_{\rm{f}}$) 
in order to find the best compromise between computational load and search sensitivity. 
As an example, in Fig. \ref{fig:Step_vs_tfft} we plot the frequency grid step $\Delta_{\rm{f}}$ as a function of the FFT duration, $T_{\rm FFT}$, for a source direction $(\lambda=193.3162^o,~\beta=-30.9956^o)$. 
Increasing $T_{\rm FFT}$ improves the sensitivity, at the same time requiring a smaller step $\Delta_{\rm{f}}$, hence a larger number of frequency grid points over $B_0$ to search for.
Moreover, in the more general situation in which position and spin-down are not exactly known, the number of grid points in the parameter space increases proportionally to $T^4_{\rm FFT}$ \cite{AsDaFrPa:2014_PRD}, further increasing the computational cost of the analysis.

For implementation purposes, see Sec. \ref{implement}, $T_{\rm FFT}$ is taken as an integer number of sidereal days, $T_\oplus =86164.09$ seconds.

\section{5-vector Statistics}
\label{sec:fivevec}
After the heterodyne coarse correction of the Doppler effect, a CW signal is still not monochromatic in each data segment, if $T_{\rm FFT}$ is larger than one sidereal day. This is due to the sidereal modulation induced by the time-varying detector response toward the source direction. In this section we remind the definition of the \textit{5-vector} statistics, which allows to take this effect into account.

The signal at the detector, after the Doppler correction, is given by \cite{2010CQGra..27s4016A}
\begin{equation}
h(t)=Re\left[H_0(H_+ A_+(t) + H_\times A_\times(t))e^{i(\omega_0 t + \phi_0)}\right]\,
\label{eq:gwsignal}
\end{equation}
where $H_0$ is the signal amplitude, and
\begin{equation}
\begin{aligned}
H_+=\frac{\cos{2\psi}-i \eta \sin{2\psi}}{\sqrt{1+\eta^2}}\\
H_\times=\frac{\sin{2\psi}+i \eta \cos{2\psi}}{\sqrt{1+\eta^2}}.
\end{aligned}
\label{eq:plustimes}
\end{equation}
In Eq. \ref{eq:plustimes} $\eta=\frac{-2\cos{\iota}}{1+\cos^2{\iota}}$, with $\iota$ the angle between the source rotation axis and the line of sight, while $\psi$ being the wave polarization angle. The two functions
$A_{+/\times}(t)$ are periodic functions of the Earth sidereal angular frequency $\Omega_\oplus=2\pi/T_\oplus$. They are linked to the classical radiation pattern functions $F_{+/\times}(\psi;t)$ \cite{Jaranowski:1998qm} by $A_{+/\times}=F_{+/\times}(\psi=0)$. The signal amplitude $H_0$ in Eq. \ref{eq:gwsignal} is related to the classical strain amplitude $h_0$ by the relation
\begin{equation}
h_0=\frac{2H_0}{\sqrt{1+6\cos^2{\iota}+\cos^4{\iota}}}.
\label{eq:h0conv}
\end{equation}
The sidereal modulation, which affects both the amplitude and the phase of the signal, produces a splitting of the signal power in five frequencies, $\omega_0\pm \textbf{k} \Omega_\oplus,~\textbf{k}=0,\pm 1,\pm 2$. This was exploited in \cite{2010CQGra..27s4016A} to introduce a detection statistics, based on the concept of \textit{5-vector}, defined as the complex vector containing the Fourier components of the signal at the five frequencies associated to the sidereal modulation. The \textit{5-vector} of a given time series of duration $T_{\rm FFT}$, and at an angular frequency $\omega_\mathrm{0}$, is given by (working, for simplicity of notation, in the continuous)
\begin{equation}
\label{5vec}
 \textbf{X} =\int_{T_{\rm FFT}}x(t) e^{-i\left(\textbf{k} \Omega_\oplus + \omega_0\right) t}dt.
\end{equation}
In addition to the data \textit{5-vector} $\textbf{X}$, the signal template \textit{5-vectors} $\textbf{A}_+, \textbf{A}_\times$ are also computed, for each $\omega_\mathrm{0}$, by means of Eq. \ref{5vec}, replacing the time series $x(t)$ with the two functions $A_{+/\times}(t)$. Although analytical formulae have been derived for $A_{+/\times}(t)$, see e.g. \cite{2010CQGra..27s4016A}, the corresponding \textit{5-vectors} are computed numerically in order to take into account features of real data, for instance data gaps, which would be difficult to deal with otherwise. These three \textit{5-vectors} are then combined computing two matched filters of the data with the signal templates:
\begin{equation}
\label{estim}
\hat{H}_+=\frac{\textbf{X}\cdot \textbf{A}^+}{|\textbf{A}^+|^2} \qquad \text{and} \qquad \hat{H}_\times=\frac{\textbf{X}\cdot \textbf{A}^\times}{|\textbf{A}^\times|^2}.
\end{equation}
It can be shown \cite{2010CQGra..27s4016A} 
that $\hat{H}_{+/\times}$ are estimators of the quantities  $H_\mathrm{0}H_{+/\times}e^{i\phi_0}$ in Eq. \ref{eq:gwsignal}. They are used to define the \textit{5-vector } statistics as
\begin{equation}
\mathcal{S}=|\textbf{A}_+|^4 |\hat{H}_+|^2 +|\textbf{A}_\times|^4 |\hat{H}_\times|^2,
\label{5vectdetstat}
\end{equation}
which collects the signal power, spread due to the sidereal modulation, over a time interval $T_{\rm FFT}$, and which also depends on the detector noise through the data \textit{5-vector}.
\section{Removal of the residual Doppler and computation of the final statistics}
\label{sec:semicoh}
In principle, once we have computed the \textit{5-vector} statistics for all frequencies of the grid and over all segments of duration $T_{\rm FFT}$, the final statistics value would be simply obtained by summing all the \textit{5-vector} statistics values at fixed frequency. In practice, however, we have to take into account the remaining frequency spread due to the coarse Doppler correction described in Sec. \ref{sec:tfft}, otherwise the signal power at a given frequency would not be fully recovered. 

\begin{figure}
    \includegraphics[width=\columnwidth]{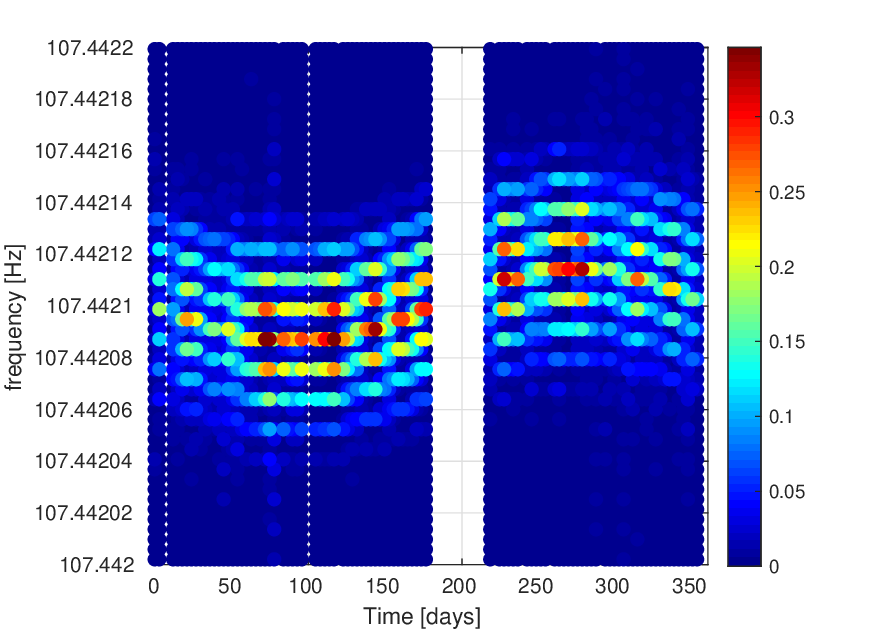}
    \includegraphics[width=\columnwidth]{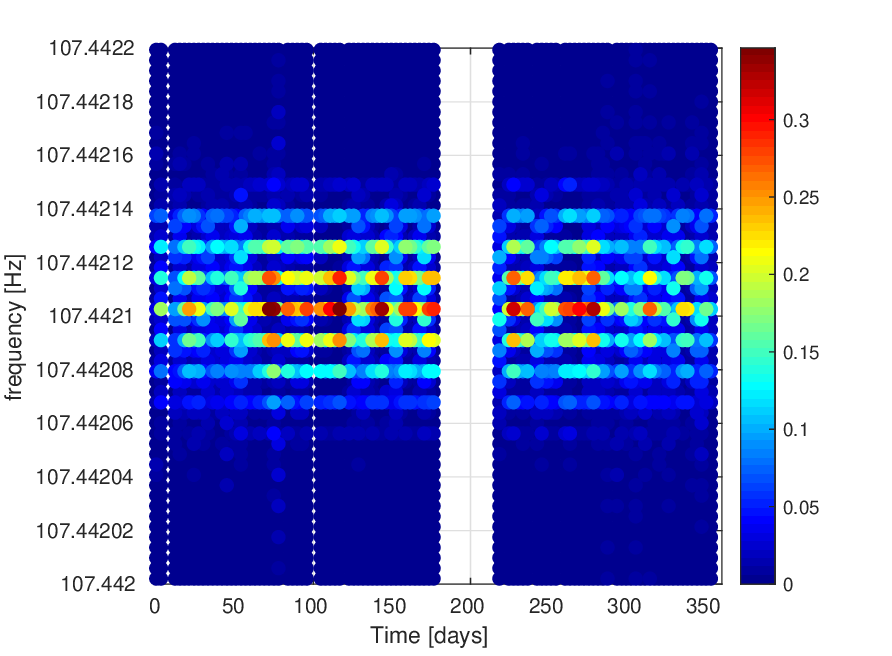}
    \caption{Time-frequency plot of the statistics computed with $T_{\rm FFT}=3 T_\oplus$ for a simulated source with unitary amplitude, sky position and polarization parameters corresponding to signal $s_1$ in Tab. \ref{tab:TAB_P3_P5_pol_pos}, frequency $=107.4421$ Hz, zero spin-down and assuming it is observed by the LIGO Livingston detector. Top plot refers to the coarse correction done for the grid frequency value nearest to the true signal frequency, while the bottom plot to the corresponding refined correction. The color bar gives the value of the 5-vector statistics computed through Eq. \ref{5vectdetstat}. Even though the plots have been obtained considering only the signal, i.e. without detector noise, data gaps of the O3 run have been taken into account in the simulation. }
    \label{fig:Stat_unc_and_cor_DOPPLER}
\end{figure}

The choice of $\Delta_{\rm{f}}$, for a given $T_{\rm FFT}$, on the basis of Eq.\ref{eq:condition} guarantees that, in each time interval $T_{\rm FFT}$, the signal power remains confined into a frequency bin. As a consequence, in each time segment the \textit{5-vector} statistics is not affected by the not optimal Doppler correction.
The residual Doppler, however, acts by off-setting the values of the statistics in different segments of the time-frequency plan, as can be clearly seen in the top plot of Fig. \ref{fig:Stat_unc_and_cor_DOPPLER}, obtained considering a simulated signal of unitary amplitude with parameters shown in the first row of Tab. \ref{tab:TAB_P3_P5_pol_pos} (``$s_1$''), generated assuming it is observed by the LIGO Livingston detector. The plot shows the time-frequency map of the \textit{5-vector} statistics of the signal after the \textit{coarse} Doppler correction, for the coarse frequency value nearest to the true signal frequency, taking data segments of duration $T_{\rm FFT}=3 T_\oplus$. In this case the residual Doppler amounts to about $8\times 10^{-5}$ Hz, which is much smaller than the full uncorrected Doppler shift, $\simeq 0.01$ Hz, but larger than the frequency bin of $\frac{1}{3 T_\oplus} \simeq 3.9\times 10^{-6}$ Hz. 
Therefore, before summing the statistics on the time axis, to obtain the final semi-coherent statistics, we need to properly shift the frequencies in order to realign them correctly.
Specifically, from Eq. \ref{eq: f_variation} it follows that
\begin{equation}
\label{eq:f_0}
f_0=\frac{f(t)+f'_\mathrm{0}\frac{\vec{v}_{\rm t} \cdot \hat{n}}{c}}{(1+\frac{\vec{v}_{\rm t} \cdot \hat{n}}{c})},
\end{equation}
where $f'_\mathrm{0}$ denotes the frequency grid values.
We have to shift the frequencies $f(t)$ by an amount $D_f(t)$ such that $f(t)-D_f(t)=f_\mathrm{0}$.
It thus follows, from Eq. \ref{eq:f_0}, that
\begin{equation}
\label{eq:DF}
D_f(t)=\frac{(f(t)-f'_\mathrm{0})\frac{\vec{v}_{\rm t}\cdot \hat{n}}{c}}{(1+\frac{\vec{v}_{\rm t} \cdot \hat{n}}{c})},
\end{equation}
and the new corrected frequencies are obtained as
\begin{equation}
\label{eq:SHIFT}
f_c(t)= f(t)-D_f(t)
\end{equation}
By construction, this shift will re-align the signal peaks only when $f'_\mathrm{0}$ is the grid value nearest to the true signal frequency. 
The bottom plot of Fig. \ref{fig:Stat_unc_and_cor_DOPPLER} shows the time-frequency plot of the \textit{5-vector} statistics, after the \textit{refined} Doppler correction, considering the nearest point to the signal frequency. As expected, after the refined correction the statistics peaks are aligned. Due to the computation of scalar products between the signal and the templates at different frequency bins (Eq. \ref{estim}), the statistics presents 9 prominent peaks, although both the signal and the templates are characterized by only 5 peaks. See Sec. \ref{sec:spindown} for a more detailed discussion.
In the simulation, we have taken into account gaps in O3 Livingston data: the empty region in the plots corresponds to a 1-month detector commissioning break that occurred during the run. 
\begin{figure}[htb]
    \centering
    \includegraphics[width=\columnwidth]{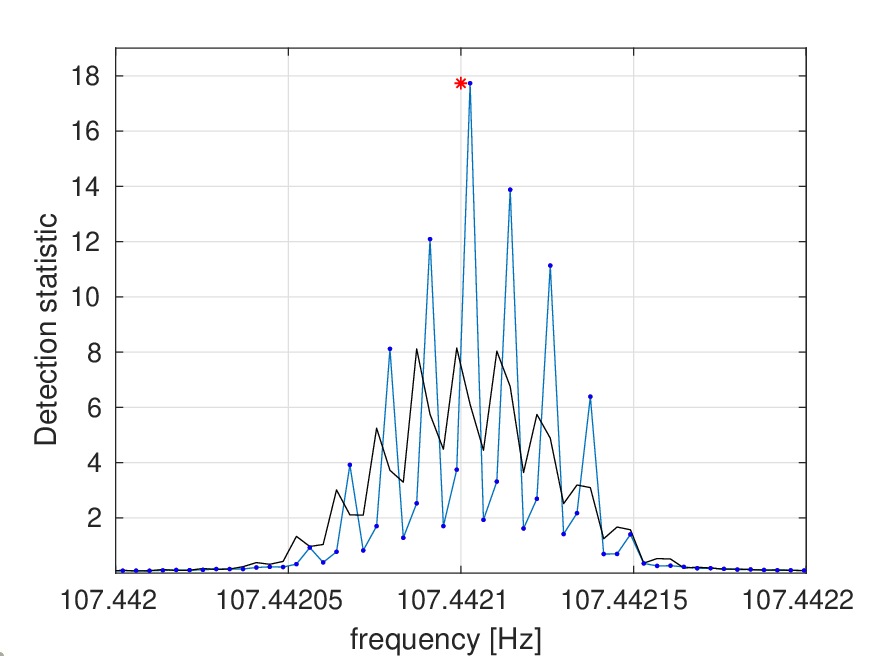}
     \caption{Global statistics, see Eq. \ref{eq:stattot},  before (black, continuous line) and after (blue, dotted line) the refined correction for a simulated signal with parameters given in the caption of Fig. \ref{fig:Stat_unc_and_cor_DOPPLER} and assuming it is observed by the LIGO Livingston detector. The asterisk indicates the signal frequency.}
    \label{fig:SUM_stat}
\end{figure}
Fig. \ref{fig:SUM_stat} shows the final statistics before and after the residual Doppler correction. Again, the effect of the correction is clearly visible and produces stronger peaks. When the refined step is applied to the other coarse grid points, the resulting correction will be less accurate and produce less significant peaks in the final statistics.

\section{Spin-down correction}
\label{sec:spindown}
\begin{figure}
    \centering
    \includegraphics[width=\columnwidth]{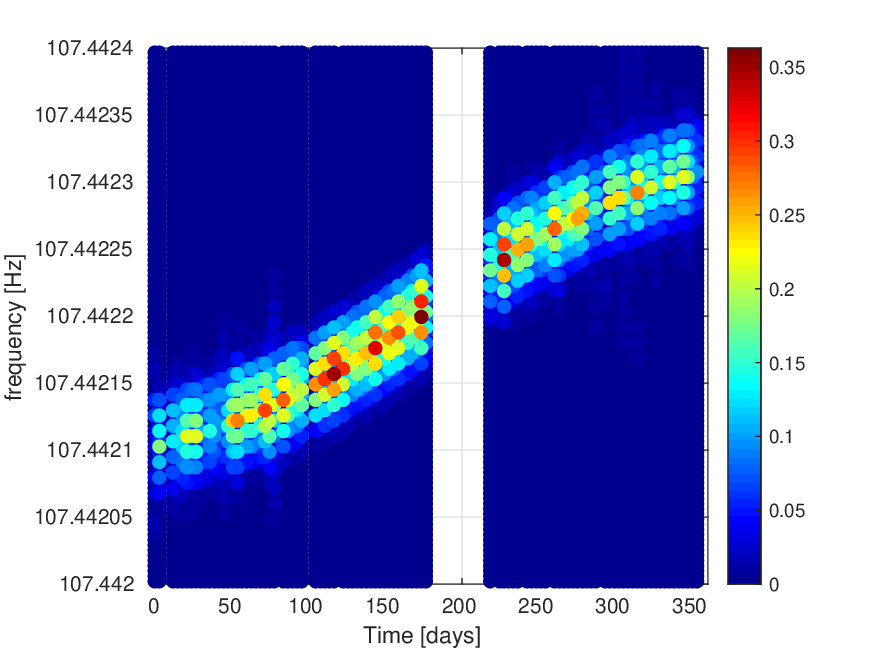}
    \includegraphics[width=\columnwidth]{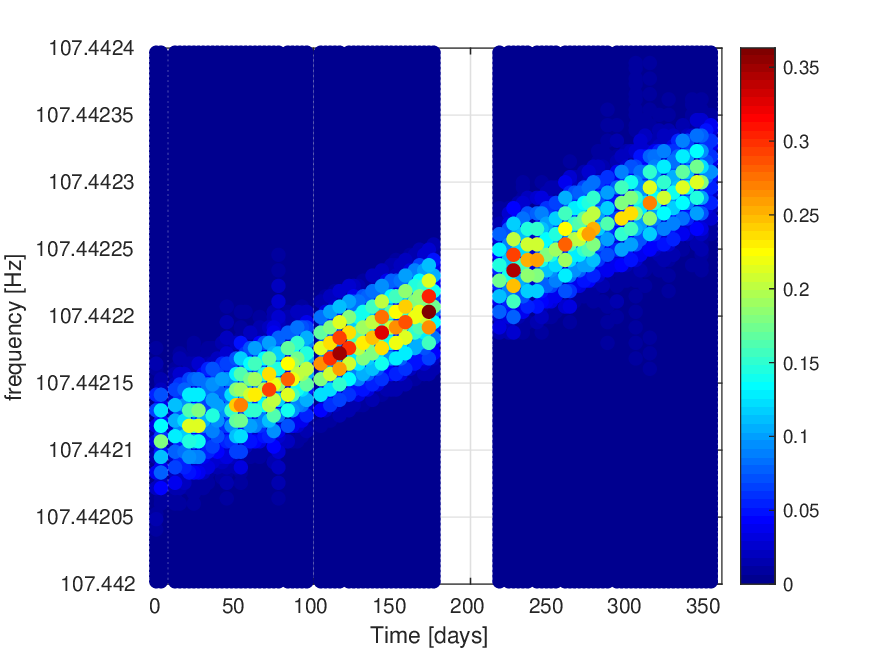}
    \includegraphics[width=\columnwidth]{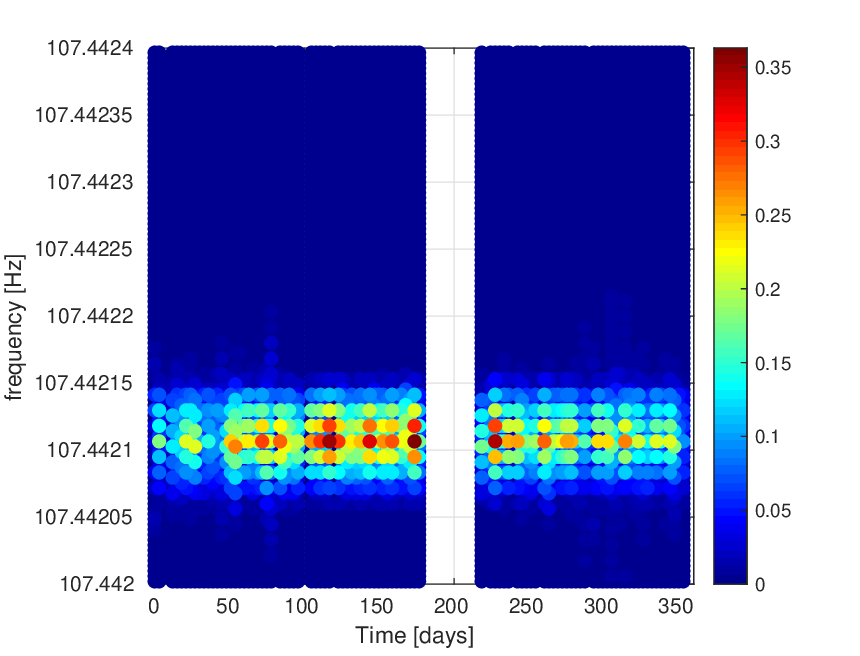}
    \caption{Time-frequency plot of the detection statistics for a simulated signal with unitary amplitude, parameters given in the text and assuming it is observed by the LIGO Livingston detector. The analysis has been carried on with $T_{\rm FFT}=3 T_\oplus$. Top plot: statistics after Doppler and spin-down coarse correction, using the frequency and spin-down values nearest to the correct signal values. Middle plot: statistics after refined Doppler correction. Bottom plot: statistics after refined spin-down correction, done using the refined spin-down value nearest to the correct one. Even though the plots have been obtained considering only the signal, i.e. without detector noise, data gaps of the O3 run have been taken into account in the simulation.}
    \label{fig:DOP_SD_COR}
\end{figure}

\begin{figure}
    \centering
\includegraphics[width=\columnwidth]{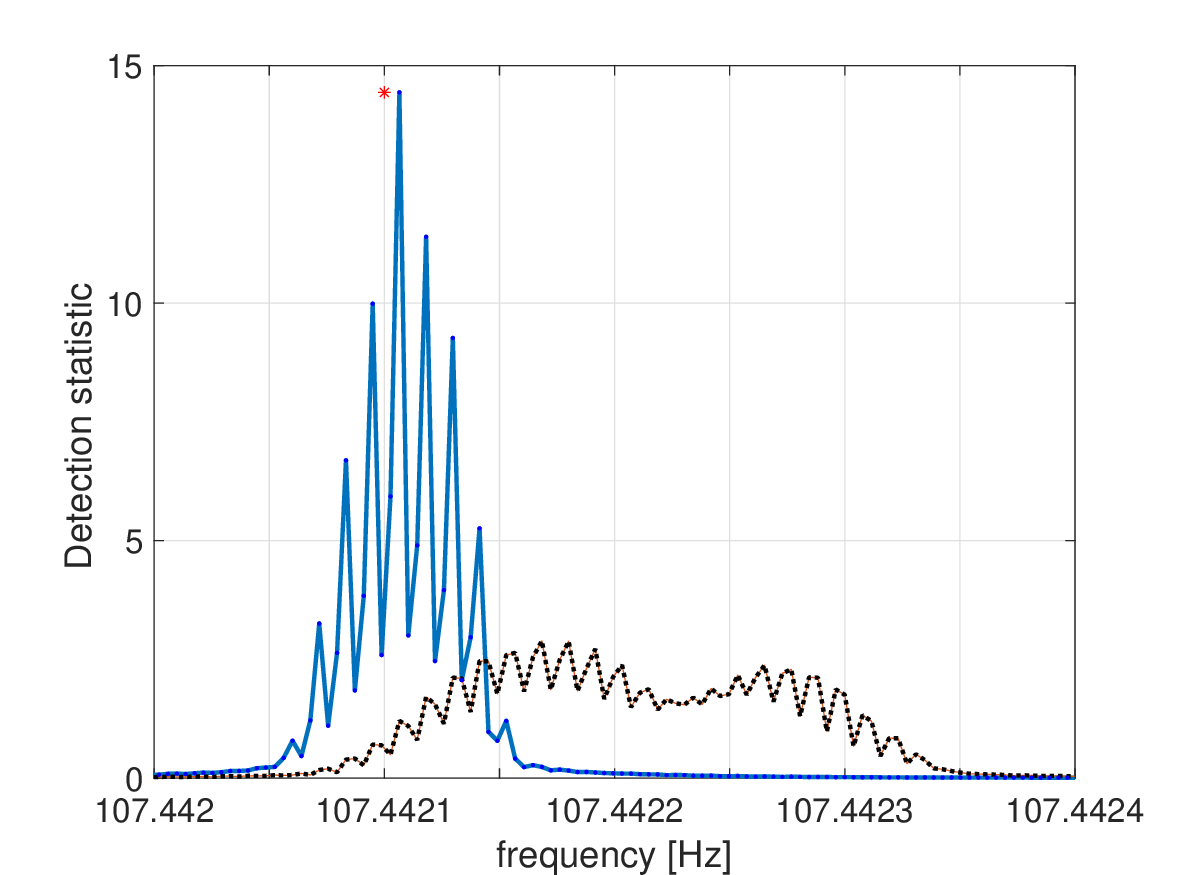}
  \caption{Final statistics, see Eq. \ref{eq:stattot}, as a function of the frequency, before (red, dotted line) and after (blue, continuous line) spin-down correction for $T_{\rm {FFT}}=3 T_\oplus$ for a simulated signal of unitary amplitude with parameters given in the text. The star indicates the signal frequency.}
    \label{fig:DOP_SD_COR_Proj}
\end{figure}

In previous sections we have focused on the Doppler correction, neglecting any intrinsic frequency variation of the signal.
In this section, we describe how to remove from the data the frequency variations due to the spin-down. In particular, we consider here only the correction for the first order spin-down term (i.e the first time-derivative of the frequency). As discussed in Appendix \ref{app:secondsd}, for any fixed $T_{\rm FFT}$ longer than one sidereal day, some portions of the potentially explorable parameter space - especially for very large absolute values of the first order spin-down - would require the second spin-down term is also taken into account. The correction of the second order spin-down is deferred to a future work.
Suppose we carry out a search for a source located at a given sky position and emitting a CW signal with unknown frequency and spin-down, ranging respectively in the interval $[{f_{\rm min}}, ~{f_{\rm max}}]$ and $[\dot {f}_{\rm min},~\dot{f}_{\rm max}]$. As for the Doppler, also in this case we first apply a \textit{coarse} spin-down correction followed by a \textit{refined} correction to substract the residual spin-down frequency variation and to re-align the frequencies of the statistics values due to a CW signal.
For a fixed coherence time $T_{\rm FFT}$, the \textit{coarse} spin-down step, defined as the maximum mismatch such that the signal power, during the time interval $T_{\rm FFT}$, is confined to a single frequency bin, is 
\begin{equation}
    \delta \dot{f_0} = \frac{\delta f}{2T_{\rm FFT}}.
    \label{eq:dfdot0}
\end{equation}
For that given sky position, we then perform a coarse heterodyne data correction (for the Earth Doppler effect), to scan the frequency range of interest at steps $\Delta_f$ as discussed in Sec. \ref{sec:tfft} and, for each frequency, a coarse heterodyne spin-down correction at spin-down values  $\dot{f_{\rm n}}=\dot{f}_{\rm min}+n\times \delta \dot{f_0}$, where $ n=0, ...\mathrm{round}(|\dot{f}_{\rm max}|/ \delta \dot{f}_0)$. After this stage, the time frequency-plot of the statistics is affected by the residual Doppler and spin-down effects, due to non-optimal signal corrections, that propagate over the observation time as shown in the top plot of Fig. \ref{fig:DOP_SD_COR}, which refer to the values of the coarse frequency and spin-down nearest to the true signal values.
In this example we show, for illustrative purposes, the results of the analysis performed on a fake signal, with unitary amplitude, emitted by a source with sky position and polarization parameters corresponding to signal $s_1$ in Tab. \ref{tab:TAB_P3_P5_pol_pos}, $f=107.4421~Hz$, $ \dot{f}=-8.34~ 10^{-11} \rm {Hz/s} $, assuming it is observed by the LIGO Livingston detector.
We have run the algorithm over the band $[107-108]$ Hz, with step $\Delta_{\rm{f}}=0.3275$ Hz, as given by Eq. \ref{eq:condition} for the specific value of $T_{\rm{FFT}}=3T_\oplus$.
At this point, we remove the residual Doppler due to the approximate frequency correction by properly shifting the frequency of the statistics values, see Eq. \ref{eq:SHIFT}. The middle plot of Fig. \ref{fig:DOP_SD_COR} shows that the signal, after the residual Doppler removal, is only affected by the residual spin-down, due to the previous not optimal spin-down correction.
Now we make a loop over the \textit{refined} spin-down grid, with step
$\delta \dot{f} = \frac{\delta f}{2\rm T_{obs}}$, which now covers the interval between each pair of successive spin-down values of the coarse grid. 
The grid step is chosen in such a way that the whole signal power, over the total observation time $\rm T_{obs}$, is confined within a single frequency bin.
In practice, the correction is performed by shifting the frequency of the statistics values according to the rule
\begin{equation}
\label{eq:SHIFT_SD}
f_c(t)= f(t)-K\delta \dot{f} \cdot t ,
\end{equation}
where $K=1,.., \mathrm{round}(\delta \dot{f}_0/ \delta \dot{f})-1$, and the time $t$ refers to the central time of each data segment. The bottom plot of Fig. \ref{fig:DOP_SD_COR} shows the result of the refined corrections for both Doppler and spin-down effects, done using the refined spin-down value nearest to the correct one: the time-frequency values due to the signal are now aligned in frequency.
Finally, the statistics are added on the time axis. Fig. \ref{fig:DOP_SD_COR_Proj} shows the cumulative corrected detection statistics before and after the spin-down correction.
We find that both the frequency and the spin-down of the fake signal have been correctly recovered inside the refined frequency and the refined spin-down bins.

As already noticed in Sec. \ref{sec:semicoh}, from Figs. \ref{fig:DOP_SD_COR} and \ref{fig:DOP_SD_COR_Proj} it can be seen that actually there are nine frequency values associated to the injected signal, because the convolution between the 5-comb of the data and the 5-comb of the theoretical kernel leads to a 9-comb. The nine peaks of the statistics are separated by one sidereal frequency bin, $1/T_\oplus$, and their relative amplitude depends on the unknown signal polarization. For the specific case shown in Figs. \ref{fig:DOP_SD_COR} and \ref{fig:DOP_SD_COR_Proj}, the central peak is the most significant.
In general, the effect of signal polarization, combined with that of noise fluctuations, can result in a most prominent peak different from the central one, that is, not corresponding to the real signal frequency.
This has implications for the choice of the coincidence window, as we show in Sec. \ref{sec:chara}, when outliers found in different datasets are compared.

\subsection{Summary of the method}
We conclude this section with a brief summary of Sec. \ref{sec:semicoh} and \ref{sec:spindown} in order to clarify the main steps of the pipeline.
Consider a data set covering a frequency band of width $B$. For a chosen coherence time $T_{\rm{FFT}}$, and sky position, a coarse frequency and spin-down grid are defined.
For each point of the coarse frequency grid, with step $\Delta_f$ derived from Eq. \ref{eq:condition}, and for each point of the coarse spin-down grid, with step $\delta \dot{f}_0$ given by Eq. \ref{eq:dfdot0}, a coherent data correction over the whole dataset is performed via heterodyne. A time-frequency map of the detection statistics is computed over data segments of length $T_{\rm{FFT}}$, through the definition in Eq. \ref{5vectdetstat}.
For each coarse frequency value, the residual Doppler is removed by Eq. \ref{eq:SHIFT} to get the time-frequency Doppler-corrected statistics.
On the corrected time-frequency map, further shifts are applied for the refined spin-down values between each pair of consecutive coarse spin-down bins, via Eq.\ref{eq:SHIFT_SD}. Finally, for each refined spin-down value, the sum of the statistics $\mathcal{S}_{\rm i}$ for each time segment of duration $T_\mathrm{FFT}$ is computed , where the index $i$ identifies the frequency bin, of width $1/T_{\rm{FFT}}$. The final semi-coherent statistics is then a function of the frequency
\begin{equation}
    \mathcal{S}(\mathrm{f}) = \sum_{{\rm k}=1}^{\rm N} \mathcal{S}_{\rm k},
    \label{eq:stattot}
\end{equation}
where the sum extends to the  $\mathrm{N}=\mathrm{floor}\left(\frac{T_\mathrm{obs}}{T_\mathrm{FFT}}\right)$ segments contained in the observation window. 
Outliers are selected on this final statistics, dividing the searched frequency band in a number of subbands, and choosing a given number of the most significant candidates, based on the Critical Ratio (see Sec. \ref{sec:chara} for definition). Outliers found in different datasets are then subject to further analysis steps, as briefly discussed in Sec. \ref{sec:scheme}. The whole procedure can be reapeted considering different sky positions, if needed. 

\section{PIPELINE CHARACTERIZATION}
\label{sec:chara}
This section is dedicated to characterizing the pipeline in terms of sensitivity and computational cost. We start, however, by providing a couple of implementation details which will be relevant also for real searches. 

\subsection{Implementation details}
\label{implement}
As anticipated in Sec. \ref{sec:tfft}, the data segment duration, $T_{\rm{FFT}}$, is chosen to be a multiple integer of the  sidereal day of the Earth, $T_\oplus$.
In this way the 5-vector components correspond to integer frequency bins. Hence, any 5-vector can be computed by selecting the proper frequency bins in a FFT of the data. This approach, first introduced in \cite{2017CQGra..34m5007M}, brings a significant speed-up (about three orders of magnitude) with respect to the computation based on the direct application of Eq. \ref{5vec}. 

A second detail concerns the frequency discretization that can lead to losses in the recovered signal power up to $36\%$, due to the mismatch between the signal frequency and the central frequency of the bins \cite{2002AJ....124.1788R}.
A cheap method to reduce this effect consists in estimating the FFT values at half bins, using an ``interbinning'' interpolation \cite{2002AJ....124.1788R} \cite{2017CQGra..34m5007M}:
\begin{equation}
\label{eq:interB}
 X_{FFT,k+1/2}\approx \frac{\pi}{4}(X_{FFT,k} - X_{FFT,k+1})
\end{equation}
where  $X_{FFT,k}$ denotes the value of the $FFT$ sample at the $k$-th frequency bin. The impact of interbinning in the sensitivity estimation will be discussed in the next section.

\subsection{Sensitivity}
\label{sec:sensi}
The sensitivity is defined as the minimum strain amplitude detectable with a given confidence level (C.L.), which we choose to be $95\%$. 

We have made an empirical estimation of the sensitivity via software injections of simulated signals in a few frequency bands of real detector data, which has been then extrapolated to the full frequency band 10-2048 Hz, as outlined in the following. 
We have used O3 LIGO Livingston data in three different 1-Hz frequency bands: $[107,~108]$ Hz, $[585,~586]$ Hz, $[883,~884]$ Hz.
In each of them, two sets of 80 and 40 signals, denoted as $\rm{s_1}$ and $\rm{s_2}$, have been generated, respectively with random frequency, while spin-down, position and polarization parameters were fixed at the values given in Tab. \ref{tab:TAB_P3_P5_pol_pos}, and added to the detector data.
\begin{table*}[t]
    \begin{tabular}{c|c|c|c|c}
    \toprule
  Signal & $\lambda$ [deg]  & $\beta$ [deg] & $\cos \iota$ & $\psi$ [deg] \\ \hline
  $\rm{s_1}$ & 193.3162 & -30.9956 & -0.081 & 25.4390 \\
  $\rm{s_{2}}$ & 276.8964 & -61.1909 & 0.463 &-20.8530   \\ 
    \hline
    \end{tabular}
  \caption{Position (in ecliptical coordinates) and polarization parameters of the two sets of simulated signals used to test the pipeline performances. 
  } 
    \label{tab:TAB_P3_P5_pol_pos}
\end{table*}
Each set of 80 and 40 signals has been injected from 10 to 15 times, each time with different values for the amplitude $H_0$ (see Eq. \ref{eq:gwsignal}), chosen in a range that is expected to contain the minimum detectable value, at the 95$\%$ confidence level. Data has been analyzed as they would be in a real search, considering the whole 1-Hz band but only two coarse spin-down bins\footnote{Out of $M=\mathrm{round}(|\dot{f}_{\rm max}|/ \delta \dot{f}_0)$, see discussion in Sec. \ref{sec:tfft}.} around the injected values (to save computing time). Both the coarse and the refined corrections have been applied. 

\begin{figure}
    \centering
    \includegraphics[width=\columnwidth]  {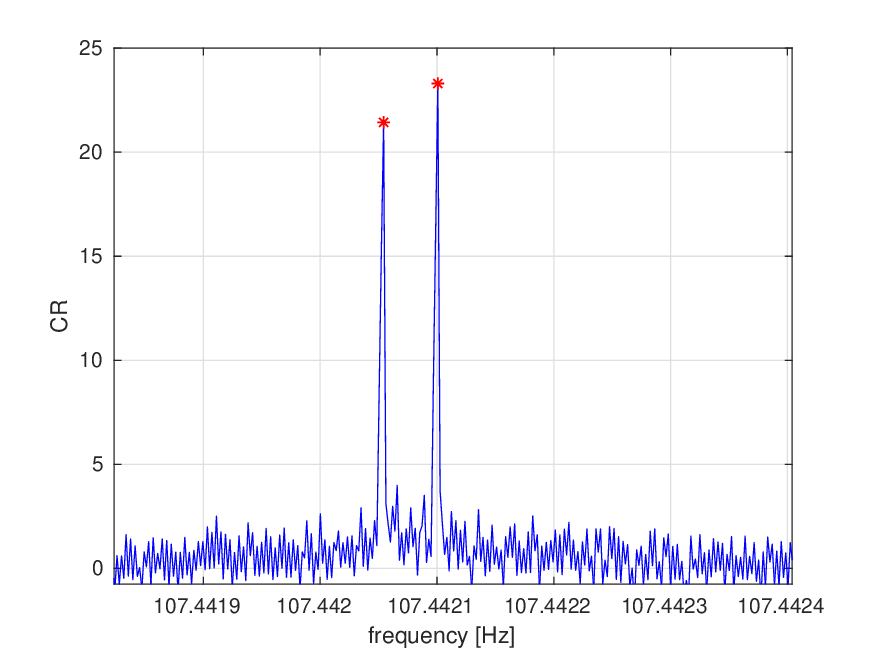}
    \caption{Critical Ratio $\mathcal{CR}$ over a frequency band containing the simulated signal $\rm{s_2}$  see Tab. \ref{tab:TAB_P3_P5_pol_pos}, injected in O3 LIGO Livingston data and amplitude $H_0=10^{-25}$. The statistics shows two peaks having comparable significance. The two peaks are distant $4$ sidereal frequency bins, with the smaller one at the signal frequency.}
    \label{fig:freqError_signal}
\end{figure}

\begin{figure}
    \centering
    \includegraphics[width=\columnwidth]{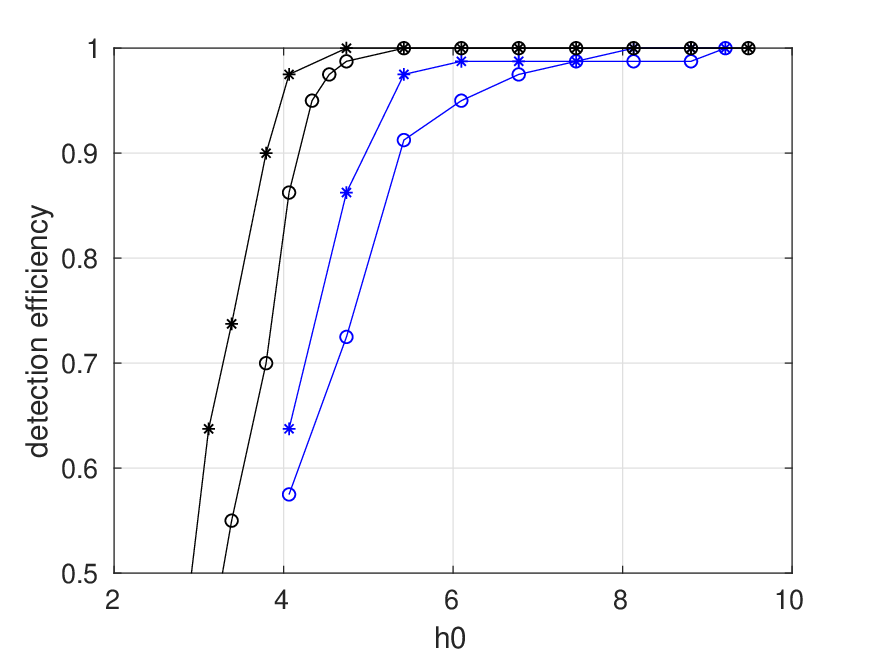}
    \caption{Detection efficiency curves, done using several sets of $80$ software simulated signals with the parameters of the HI P3, injected in the band [107-108] Hz of LIGO Livingston O3 data with different amplitudes. The pair of grey curves are the detection efficiencies, as a function of the signal amplitude, without interbinning, and  
    using a coherence time of $5$ (circles) and $10$ (asterisks) sidereal days.
    The pair of black curves correspond to the application of the interbinning procedure, which significantly increases the detection efficiency. Signal amplitude is in units of $10^{-26}$}.
    \label{fig:INTERP_COMPAR}
\end{figure}

\begin{figure*}
    \centering
\includegraphics[width=\textwidth]{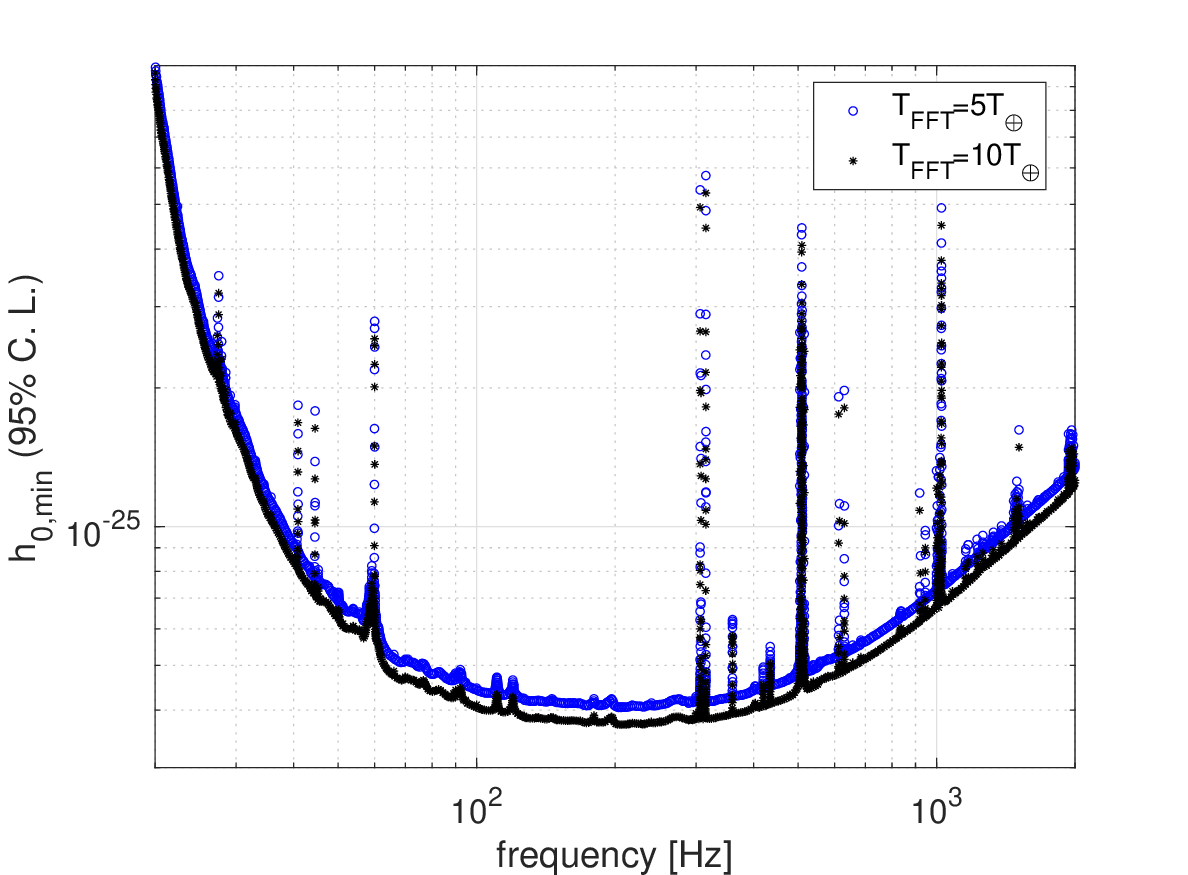}
    \caption{Search estimated sensitivity for O3 LIGO Livingston (95$\%$ C.L.), obtained through the injection of simulated signals in detector real data. The curves correspond to two different segment duration, $5T_\oplus$ (grey, circles) and $10T_\oplus$ (black, asterisks). Interbinning has been applied, see text for more details.}
    \label{fig:SENS1}
\end{figure*}

For each set of injected signals of amplitude $H_0$, and for each spin-down value, after running the analysis we select the 300 most significant outliers, accross the 1-Hz frequency band.
As standard in several wide-parameter searches, the significance of a candidate is represented by its Critical Ratio CR \cite{AsDaFrPa:2014_PRD} computed on the projection on the frequency axis of the time-frequency map of the detection statistics values, and defined as 
\begin{equation}
\rm CR(f)=\frac{\mathcal{S}(f)-\mu_{n}}{\sigma_{n}},
\end{equation}
where $\mathcal{S}$ is given by Eq. \ref{eq:stattot}, 
$\mu_{n}$ and $\sigma_{n}$ are the mean and standard deviation of the noise statistics. 
The noise statistics is evaluated by replacing the data five-vector $\textbf{X}$, Eq. \ref{estim}, with a noise five-vector whose components are randomly chosen over the one Hz frequency band so that it cannot represent a physical signal. 
Adapting the procedure typically used in searches for selecting coincidences among outliers found in different datasets \cite{2019PhRvD.100b4004A}, we choose as outliers those points in the search parameter space $(f,\dot{f})$ (the sky position is fixed), for which 
the a-dimensional distance from any of the injected signals is smaller than the coincidence window $D_\mathrm{max}$. In other words, a signal is considered as detected if the dimensionless distance \cite{AsDaFrPa:2014_PRD}
\begin{equation}
    d = \sqrt{\left(\frac{\Delta f }{\delta f \times \Delta}\right)^2+\left(\frac{\Delta \dot f}{\delta \dot f}\right)^2}\le D_\mathrm{max}
\end{equation}
where $\Delta f,~\Delta \dot f$ are the dimensional distances of the outlier from the injected signal,
and the factor $\Delta$ weights in the proper way the frequency distance that can be as large as four sidereal frequency bins ($4/T_\oplus$). This is due to the unknown source polarization, as shown in Fig.\ref{fig:freqError_signal}. The choice of $D_\mathrm{max}$ has been studied in \cite{2019PhRvD.100b4004A} in the context of all-sky searches. In practice, here we take 
$ \Delta=4$ (in units of sidereal frequency $1/T_\oplus$), and, conservatively, $D_\mathrm{max}=2$\footnote{In the case of coincidences among outliers found in different datasets, the value of $D_\mathrm{max}$ is chosen depending on the number of follow-ups that can be afforded, given an available amount of time and computing power. A bigger $D_\mathrm{max}$ allows for a more sensitive search, at the cost of a bigger number of outliers to be followed-up.}. 
The expected number of random outliers, due to noise, which verify the above condition is much smaller than one.
For each signal amplitude we count the fraction of detected signals and construct detection efficiency curves. 
Fig. \ref{fig:INTERP_COMPAR} shows an example of detection efficiency curves, for two different coherence times $T_{\rm{FFT}}=5T_\oplus$ (circles) and $T_{\rm{FFT}}=10T_\oplus$ (asterisks), with (black curves) and without (grey curves) ``interbinning''. 
The sensitivity gain is clearly visible both when we use 10 sidereal days, rather than 5, and when the interpolation is used. 
The signal amplitude $H_0$, such that $95\%$ of the injected signals have a coincident outlier corresponds to the sensitivity $H_{0,95\%}$ for a specific set of source parameters and for the specific 1-Hz frequency band we are considering. In practice, the $95\%$ level is estimated by linearly interpolating the detection efficiency between the data point immediately below and below that value.\\
In order to compute an average sensitivity on the standard strain amplitude $h_0$, we re-scale $H_{0,95\%}$ with two factors, one to average over the sky and polarization from the specific sky position and polarization angle used in the injections, and one to go from $H_{0,95\%}$ to $h_{0,95\%}$, given by the mean value of the coefficient in Eq. \ref{eq:h0conv}, which implies the average over the cosine of the star’s inclination angle. This is equivalent, and computationally much cheaper, to generating signals with random parameters.
At the end, we have three values of the sensitivity, for the three frequency bands we are considering, which correspond to regions with different detector noise. 
Each of these three sensitivity values has been extrapolated to the full frequency band
by applying a further frequency dependent scaling factor, given by the square root of the ratio
$S_n (f)/<S_{n,j}>$ of the data power spectrum estimation $S_n (f)$, over the band 10-2048 Hz, to the average power spectrum estimation (with respect to the frequency) in each of the three bands where the injections have been done, i.e. $<S_{n,j}>$, with $j=1,2,3$ corresponding, respectively to [107, 108] Hz, [585, 586] Hz, [883, 884] Hz. 
The final sensitivity estimation is given by the average (computed in each frequency bin) of the three sensitivity curves.
Fig. \ref{fig:SENS1} shows the final sensitivity $h_{\rm{min,95\%}}$ for O3 LIGO Livingston data, for two different segment durations, $T_{\rm{FFT}}=5 ~T_\oplus$ and $T_{\rm{FFT}}= 10 ~T_\oplus$. The number of combined data segments depends on $T_{\rm{FFT}}$ and on the presence of gaps in the data, and is 60 for $T_{\rm{FFT}}=5 ~T_\oplus$  and 30 $T_{\rm{FFT}}= 10 ~T_\oplus$.
From the sensitivity curves we have estimated a sensitivity depth \cite{behnke2015postprocessing,dreissigacker2018fast} $\mathcal{D}=\sqrt{S_n(f)}/h_{\rm{min,95\%}}(f)=124\pm 4$ for $T_{\rm{FFT}}= 10 ~T_\oplus$.
The sensitivity values $h_{\rm{min,95\%}}$ for specific frequency bands and for the above mentioned data segment duration $T_{\rm{FFT}}$, together with the corresponding CR, $CR_{95\%}$, are shown in Tabs. \ref{tab:SI_Sens_LL} and \ref{tab:SI_Sens_LL_P5}. The estimated sensitivity has an associated false alarm rate of $O(10^{-5})$, which corresponds to an expectation of $O(10)$ outliers in Gaussian noise for a 1-Hz frequency band, one refined spin-down bin and one sky location.  
We have also computed a theoretical sensitivity, via a mixed analytical-numerical approach, which assumes Gaussian and stationary noise, as described in Appendix \ref{theosens}. The last column of the tables reports such theoretical value. Overall, the theoretical values underestimate the empirical sensitivity, as expected due to non-Gaussian and non stationary features of real data, by at most $\sim$10-15$\%$, .    
\begin{table*}[t]
    \begin{tabular}{c|c|c|c|c|c|c}
    \toprule
   band [Hz]  & $T_{\rm{FFT}}~ [T_\oplus]$  &  $ CR_{95\%} $ & $H_{min,\rm{95\%}}$   &   $h_{min,95\%}$ &  $\sqrt{<S_n(f)>}~[Hz^{-1/2}]$& $h^{th}_{\rm{min,95\%}} $\\ \hline
  $[107, ~108]$ &  5 & 6.1 & 3.13 & 4.24 & $4.97\cdot 10^{-24}$ &3.86\\
  $[107, ~108]$ & 10 & 7.0 & 2.93 & 3.97 & & 3.57\\
  $[585,~586]$  &  5 & 6.3 & 3.85 & 5.21 & $5.87\cdot 10^{-24}$ & 4.37\\
  $[585,~586]$  & 10 & 6.4 & 3.38 & 4.58 & & 3.99\\
  $[883,~ 884]$ &  5 & 7.9 & 4.94 & 6.70 & $7.53\cdot 10^{-24}$ & 6.54\\
  $[883,~ 884]$ & 10 & 7.2 & 4.62 & 6.26 & & 5.44\\
    \hline
    \end{tabular}
  \caption{Sample results from the sensitivity estimation, using the software-simulated signal $s_1$ (see Tab. \ref{tab:TAB_P3_P5_pol_pos}). The various columns represent: the injection band (at random frequency); the data segment duration used in the analysis; the mean of the CR for the detected signals; the sensitivity estimation for $H_0$ obtained through injections; the corresponding classical strain sensitivity, the corresponding average detector noise amplitude power spectrum (which - for a given frequency band - is independent on the used segment duration), and finally the theoretical estimation. All the sensitivity estimations are in units of $10^{-26}$.}
    \label{tab:SI_Sens_LL}
\end{table*}
\begin{table*}[t]
    \begin{tabular}{c|c|c|c|c|c|c}
    \toprule
   band [Hz] & $T_{\rm{FFT}}~ [T_\oplus] $ & $CR_{95\%}$ &$ H_{min,\rm{95\%}}$  & $h_{min,\rm{95\%}}$ & $\sqrt{<S_n(f)>}~[Hz^{-1/2}]$ & $h^{th}_{min,\rm{95\%}}$\\ \hline
  $[107, ~108]$ &  5 & 4.27 & 3.37   &4.37 & $4.97\cdot 10^{-24}$ & 3.25    \\
  $[107, ~108]$ & 10 & 6.19 & 3.00   &3.88 & & 3.23  \\
    \hline
    \end{tabular}
  \caption{Same as in Tab. \ref{tab:SI_Sens_LL}, but using simulated signal $s_2$ in Tab. \ref{tab:TAB_P3_P5_pol_pos}. All the sensitivity estimations are in units of $10^{-26}$.}
    \label{tab:SI_Sens_LL_P5}
\end{table*}

\subsection{Computational cost}
\label{sec:compcost}
The semi-coherent \textit{5-vector} method is not suited - at least in its current implementation - for carrying out the follow-up of a large ($\gg 10^4$) number of candidates, as those produced in a typical all-sky search. Rather, it represents an effective method to analyze deeper a relatively small number($O(10^3-10^4)$) of significant candidates.
To give an idea of the required computational cost, to analyze one year of data from $n_\mathrm{det}$ detectors, setting $T_{\rm{FFT}}=3 T_\oplus $, a frequency band of $n_\mathrm{f}$ refined bins, a number $n_{\mathrm{sky}}$ of sky points, $n_\mathrm{sd}$ refined spin-down bins, the code takes less than $2\cdot 10^{-7} n_\mathrm{f}\cdot n_{\mathrm{sky}}\cdot n_\mathrm{sd}\cdot n_\mathrm{det}$ core-hours, corresponding to about $7\cdot 10^{-4}$ seconds per template. This would correspond to about 80 core-hours for a typical follow-up, covering say $n_{\mathrm{sky}}=9$, 0.1 Hz frequency band, $5$ coarse spin-down points and a network of three detectors. It would be then able to follow-up O(5000)candidates previously selected, in about 1$\%$ of the time needed to perform the bulk of an all-sky search. 

Another reasonable use of the procedure, both in terms of sensitivity and computational cost, concerns directed searches toward, e.g., the Galactic center or globular clusters.
Assuming, for instance, to run a directed search over one year of data of a single detector, looking for a single sky point, a frequency band of 2 kHz and exploring $10$ coarse spin-down values, would take about $1.2\cdot 10^5$ core-hours. 

The algorithm is characterized by a high level of parallelism, which can be exploited on suitable hardware devices to speed it up. Porting the code on GPUs will be the subject of a future work.

\section{TESTS WITH HARDWARE INJECTIONS}
\label{sec:applica}
\textit{Hardware injections} (HI) are simulated CW signals injected during scientific runs by directly moving detector mirrors. 
Checking the ability of an analysis pipeline to correctly recover HIs is a standard validation test for CW pipelines. In this section we present results for two different kinds of tests. The first one consists in running the analysis for specific HIs, assuming the exact sky position and spin-down values, and covering a 1 Hz range around the signal frequency, using different data segment durations. Tab. \ref{tab:TestHI_INTERP} shows the results of the analysis for three HIs present in LIGO Livingston O3 data, namely P3, P5 and P11 for three different choices of $T_{\rm{FFT}}$, as indicated in the second column. The third column gives the coarse frequency bin $\Delta_f$ used in the analysis, whose value depends on the position of the source as well as the coherence time.
The fourth column is the frequency error in detecting the signal, while the last column gives the CR. HI parameters are shown in Tab. \ref{tab:hi}.
\begin{table*}[t]
    \begin{tabular}{c|c|c|c|c|c|c}
    \toprule
  HI & frequency [Hz] & spin-down ($\dot{f}$) [Hz/s] & $(\alpha,~\delta)$ [deg] & $\cos \iota$ & $\psi$ [deg] & $H_0$   \\ \hline
  P3 & 108.857 & $-1.46\times 10^{-17}$ & (178.372,-33.437) & -0.081 & 25.455 & $6.615\times 10^{-26}$    \\
  P5 & 52.808 & $-4.03\times 10^{-18}$ & (302.627,-83.839) & 0.463 & -20.853 & $3.043\times 10^{-25}$  \\
  P11 & 31.425 & $-5.07\times 10^{-13}$ & (285.097,-58.272) & -0.329 & 23.589 & $2.045\times 10^{-25}$  \\
    \hline
    \end{tabular}
  \caption{Parameters of HIs used to test the search pipeline with LIGO Livingston O3 data. Frequency and spin-down refer to the GW signal; position is in equatorial coordinates; $cos(\iota)$ is the cosine of the angle among the source rotation axis and the line of sight; $\psi$ is the wave polarization angle and $H_0$ is the signal strain amplitude.} 
    \label{tab:hi}
\end{table*}
\begin{table*}[t]
    \begin{tabular}{c|c|c|c|c}
    \toprule
  HI & $ T_{\rm{FFT}} [T_\oplus] $  & $\Delta f~ [Hz]$ & $\delta f_{err}$ [bins]& $\rm{CR}$   \\ \hline
  P3 & 3 & 0.330 & -0.34 & 17.7    \\
  P3 & 12 & 0.025 & 0.11 & 31.0  \\
  P3 & 24 & 0.007 & 0.21 & 46.7  \\
  P5 & 3 & 0.500 & 0.12 & 94.3    \\
  P5 & 12 & 0.047 & -0.02 & 291.2  \\
  P5 & 24 & 0.012 & -0.04 & 465.8 \\
  P11 & 3 & 0.380 & 0.05 & 10.1    \\
  P11 & 12& 0.027 & 0.28 & 18.5\\
  P11 & 24 & 0.007 & 0.05 & 36.0 \\
    \hline
    \end{tabular}
  \caption{Test with HI. The various columns represent: the data segment duration $T_{\rm{FFT}}$ (in sidereal days); the coarse frequency step; the error in signal frequency recovery (in bins), and finally the recovered CR.} 
    \label{tab:TestHI_INTERP}
\end{table*}
In all cases the signal is well recovered: the error in frequency recovery is always smaller than one bin and the CR increases with $T_{\rm{FFT}}$, as expected for sufficiently strong signals.   

The second test consists in running a multi-stage analysis, in which a small parameter space volume around HIs is initially considered and explored with a given segment duration, the most significant candidate selected and then followed-up with longer segment duration. This test closely resembles what would be done in the follow-up of an outlier coming from a wide parameter search. 

\begin{figure}
    \centering
    \includegraphics[width=\columnwidth]{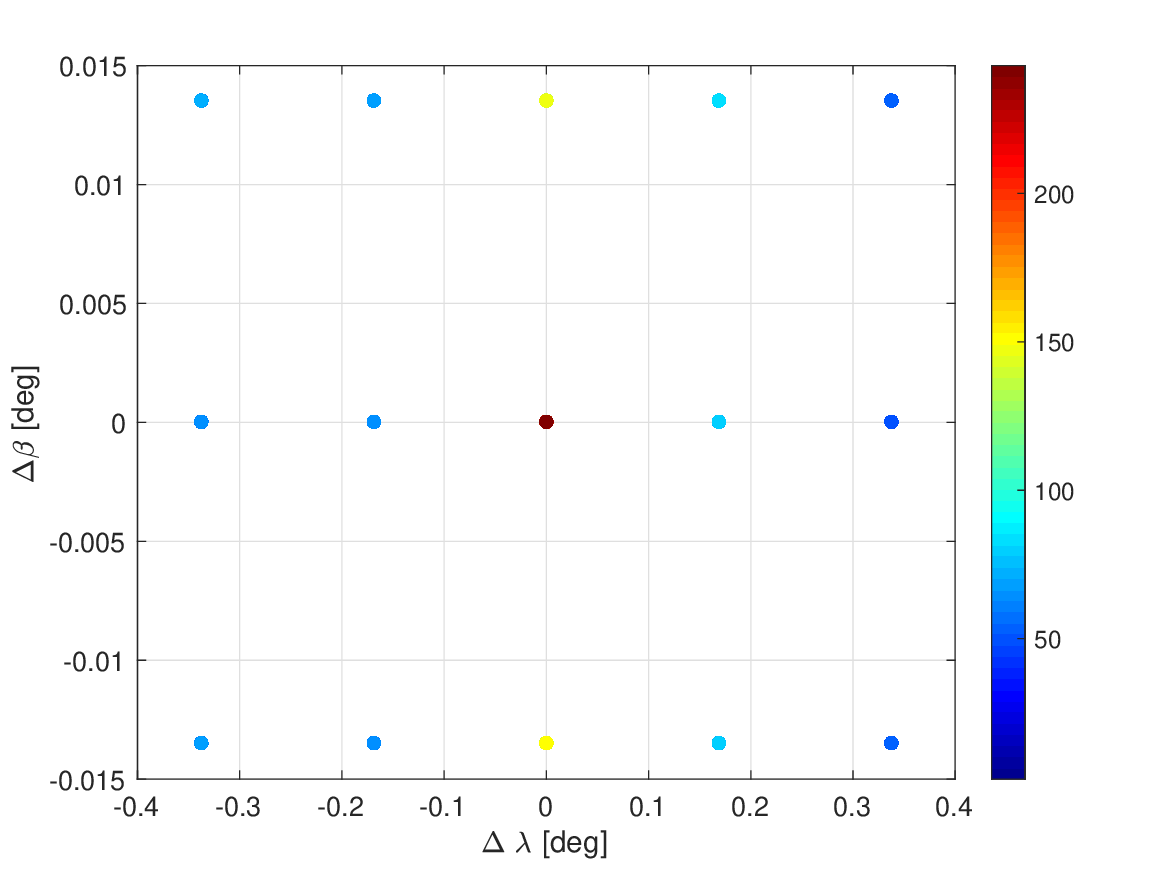}
    \caption{Analysis of the HI P5, with $T_{\rm FFT}=10T_\oplus$ (second step). We show the maximum CR over the refined sky grid around the position of the first step candidate. The maximum takes place exactly at the initial candidate position, corresponding to $\Delta\lambda=0,~\Delta\beta=0$.}
    \label{fig:FU_P5_10_SD_sky}
\end{figure}

\begin{figure}
    \centering
    \includegraphics[width=\columnwidth]{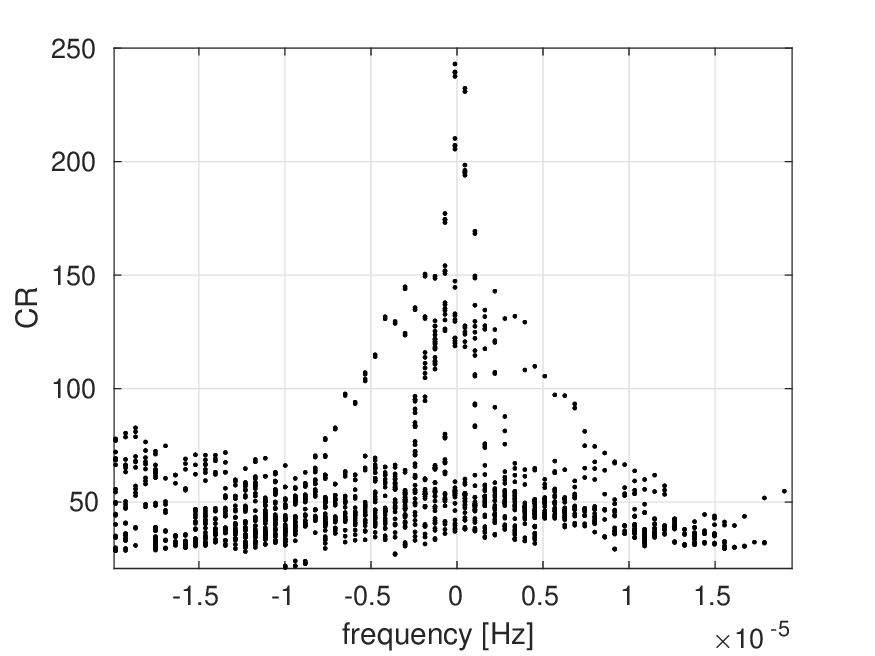}    
    \includegraphics[width=\columnwidth]{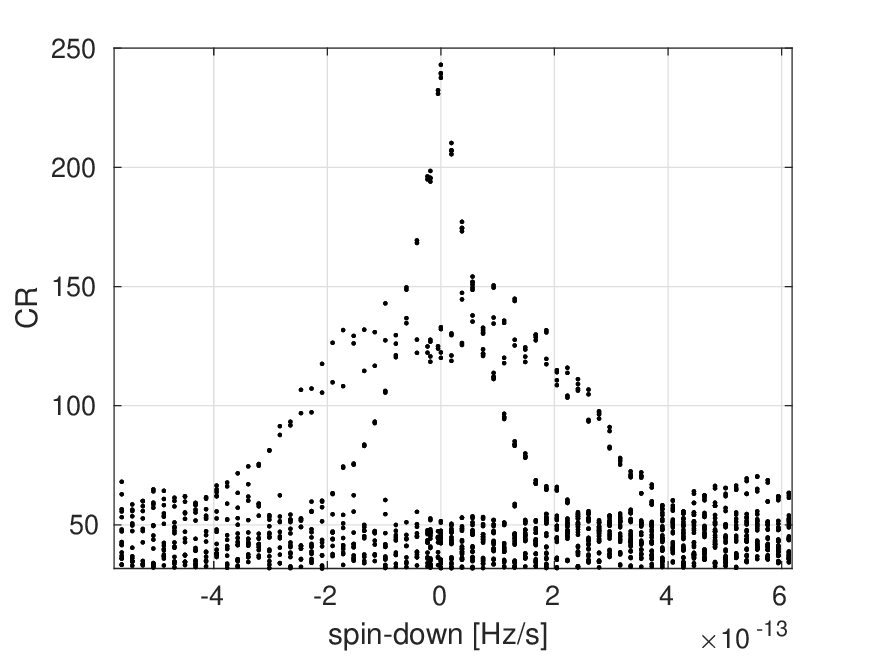}
    \caption{Analysis of HI P5, second step with $T_{\rm FFT}=10T_\oplus$: distribution of the CR as a function of the distance in frequency and spin-down from the starting candidate. The maximum takes place at a frequency and spin-down within one bin from the starting candidate values.}
    \label{fig:FU_P5_10_SD}
\end{figure}

In principle, the procedure has not any intrinsic limit on the increasing data segment duration $T_{\rm{FFT}}$, which would result in improving the sensitivity, in subsequent steps, to confirm a potential CW candidate. The maximum achievable $T_{\rm{FFT}}$ is mainly constrained by the available computing power, and is related to the size of the search parameter space, and to the number of candidates that can be reasonably handled. One further limitation may be present for nearby sources with a high transverse velocity \cite{10.1093/mnras/staa3624} with respect to the line of sight, for which using values of $T_{\rm{FFT}}$ too large would introduce a sensitivity loss due to a Doppler residual term associated to the variation of source position during the observation time. 
In the following we do not take into account this possibility, and for each HI a double step analysis has been done. First, a small portion of the parameter space, specified below, around each of source has been analyzed with a coherence time $T_{\rm{FFT}}=3 T_\oplus$. A follow-up, with $T_{\rm{FFT}}=10 T_\oplus$, has been then performed over a smaller region around the most significant candidate found in the previous step. As representative of the results, we discuss here the case of HI P5. 
The first analysis step focused on the spin-down range $[-2.2449\times 10^{-11},~ 2.2449\times 10^{-11}]$ Hz/s, a sky region covering $0.02$ deg in $\beta$ and $\pm 1.125$ deg in $\lambda$ centered at the signal position, corresponding to 15 points in the sky.
The point in this 3-dimensional grid having the highest CR ($CR\simeq$ 93) has been selected as signal candidate. It corresponds to the exact source position and to frequency and spin-down values less than one bin off the signal values. We have applied the second analysis stage in a small region around the candidate, increasing the coherence time to $10 T_\oplus$ and exploring a sky patch of $\pm 0.4$ deg  in $\lambda$ and $\pm 0.015$ deg in  $\beta$, centered at the candidate position, a frequency range of $ \pm 0.03$ Hz around the candidate values, and a spin-down range $[-6.7347\times 10^{-13},~ 6.7347\times 10^{-13}]$ Hz/s, corresponding to 4 coarse spin-down values (and 152 total refined spin-down values) around it. 
Fig. \ref{fig:FU_P5_10_SD_sky} shows the maximum CR for each point of the sky grid (whose position is measured with respect to that of the initial candidate). Fig. \ref{fig:FU_P5_10_SD} shows the distribution of the $CR$ as a function of the distance in frequency and spin-down from the starting candidate. All parameters of the most significant candidate, which has $CR=243$, are recovered within one bin of the refined grid obtained with $T_{\rm{FFT}}=10 T_\oplus$, with an error reduced by a factor of $\sim 3.3$ compared to the initial analysis.
Furthermore, the increase in CR is compatible with the improvement in sensitivity. 
Several other high CR outliers appear in Fig. \ref{fig:FU_P5_10_SD}, with slightly wrong parameters. This is due to the fact that P5 is a rather strong signal, and parameters have some degree of correlation.
\section{Conclusions}
\label{sec:concl}
In this paper we have presented a semi-coherent analysis method for the search of CW signals. The method is based on a computationally efficient incoherent combination of \textit{5-vector} statistics computed over data segments of duration larger than one sidereal day. The concept of \textit{5-vector} has been originally introduced in the context of full-coherent targeted searches and exploits the signal sidereal modulation. Here we use it as a coherent step in a semi-coherent method in which an initial coarse heterodyne Doppler and spin-down correction is followed by a more refined correction based on the shift of frequencies in the time-frequency plane. From one hand, we demonstrate the heterodyne correction is robust: in order to confine the signal power, in each time window $T_{\rm FFT}$, within the ``natural'' bin width, given by the inverse of the data segment duration, $\delta f=1/T_{\rm FFT}$, it is enough to apply such correction on a coarse grid with step $\Delta_f \gg \delta f$. On the other, we show that such coarse correction leaves a residual frequency variation which can be efficiently removed by shifting the bins of a time-frequency map built computing the \textit{5-vector} statistics over the single data segments . 

The method can be thought as a building block of a multi-step procedure in which longer and longer data segments are used, zooming in around a given interesting point (or region) in the search parameter space.  
Two natural applications are: \textit{i}) the follow-up of significant candidates, found, e.g., in all-sky searches, \textit{ii}) directed searches toward specific sky locations, like the Galactic center or globular clusters, over a large range of frequency and spin-down values.  

We have proved, by analyzing both software and hardware simulated signals injected in O3 data, that the procedure behaves as expected both in terms of improvement of the candidate significance, when the data segment duration is increased, and in terms of overall sensitivity as compared to a theoretical computation. 

The application of the method to wider parameter space, like all-sky searches, is a future milestone for which additional work is needed. 

\section*{Acknowledgement}
This material is based upon work supported by NSF's LIGO Laboratory which is a major facility fully funded by the
National Science Foundation.
In addition we acknowledge the Science and Technology Facilities Council (STFC) of the United Kingdom, the Max-Planck-Society (MPS), and the State of Niedersachsen/Germany for support of the construction of Advanced LIGO and construction and operation of the GEO600 detector. Additional support for Advanced LIGO was provided by the Australian Research Council. Virgo is funded, through the European Gravitational Observatory (EGO), by the French Centre National de Recherche Scientifique (CNRS), the Italian Istituto Nazionale di Fisica Nucleare (INFN) and the Dutch Nikhef, with contributions by institutions from Belgium, Germany, Greece, Hungary, Ireland, Japan, Monaco, Poland, Portugal, Spain.

\appendix
\section{Theoretical sensitivity}
\label{theosens}
In this Appendix we provide some details on the computation of the theoretical sensitivity we refer to in Sec. \ref{sec:sensi}. An analytical expression for the sensitivity is difficult to derive for the sum of 5-vector statistics, while it has been obtained in \cite{2014PhRvD..89f2008A} for the single 5-vector statistics, which we briefly summarize. Assuming Gaussian noise with zero mean and variance $\sigma^2$, and given the linearity of the Fourier Transform, each component of the 5-vector, defined by Eq. \ref{5vec}, is also distributed according to a Gaussian with zero mean and variance $\sigma^2_{X}=\sigma^2\cdot T_\mathrm{FFT}$. 
The two complex amplitude estimators of Eq. \ref{estim}, then, have still a Gaussian distribution with zero mean and variance $\sigma^2_{+/\times}=\frac{\sigma^2_{X}}{|\textbf{A}_{+/\times}|^2}$. As a consequence, the probability density function of the square modulus of the two estimators is an exponential and then the detection statistics defined in Eq. \ref{5vectdetstat} is distributed according to a linear combination of two exponentials with mean values $\sigma^2_{+/\times}$, see eq. 34 in \cite{2014PhRvD..89f2008A}.

In presence of a signal of amplitude $H_0$, each term of the linear combination follows a non-central $\chi^2$  distribution, with non-centrality parameter 
\begin{equation}
\beta_{+/\times}=2H_0^2|e^{j\Phi_0}H_{+/\times}\bf{A}^{+/\times}|^2/\sigma^2_{X}
\end{equation}
This applies to each term $\mathcal{S}_i$ in Eq. \ref{eq:stattot}.  
The distribution of the final statistics $\mathcal{S}$ can be numerically obtained in a straightforward way by generating the two aforementioned distributions (exponentials or non-central $\chi^2$, respectively for noise and noise plus signal) and then taking the sum in Eq. \ref{eq:stattot}. 

The theoretical sensitivity is computed in the following way. First we generate the noise-only distribution of the statistics, taking the data average power spectrum $P(f^*)$ at a given frequency, chosen a p-value $p$, e.g. 0.01, and determined the corresponding value of the statistics, $\mathcal{S}^*$. Then, for each value of the signal amplitude $H_0$, in a given range, a population of random source parameters is generated and the corresponding noise plus signal probability distribution is computed. The area of the above distribution on the right of $\mathcal{S}^*$ is evaluated, and the value $H^*_0$ such that the area equals a given value of the detection probability $\Gamma$, e.g. 0.95, is determined. The number $H^*_0$ is multiplied by a factor $\sqrt{\pi/2.4308}$, to take into account the average loss due to the uncertainty of the frequency with respect to the frequency bin center \cite{2014PhRvD..89f2008A}, and by a factor 1.3258 to convert $H_0$ to the standard strain $h_0$ (see Eq. \ref{eq:h0conv}). This number is the minimum detectable signal amplitude at C.L. $\Gamma$ and p-value $p$ for the given value of the data power spectrum. The sensitivity over the whole frequency band is obtained by multiplying that value by $\sqrt{P(f)/P(f^*)}$, i.e. the square root of the ratio of the frequency dependent data power spectrum to the reference value.   
Rather than plotting the full theoretical sensitivity, in Tab. \ref{tab:SI_Sens_LL} we reported the theoretical sensitivity computed over O3 LIGO Livingston data for a few frequency bands and different data segment durations $T_\mathrm{FFT}$, together with the empirical values obtained through the injection of simulated signals, as described in Sec. 
\ref{sec:sensi}. The two estimations are in good agreement, with the theoretical one slightly better - by 15$\%$ at the most - as expected given they are computed assuming an ideal Gaussian distribution for the noise.  

\section{Second order spin-down}
\label{app:secondsd}
The spin-down correction described in Sec. \ref{sec:spindown} regards only the first order spin-down term, $\dot{f}$. A second order term, $\ddot{f}$, would not be corrected and could determine a sensitivity loss. The condition for an uncorrected second order spin-down term to not produce a sensitivity loss is that the frequency variation it causes during the observing period, $T_\mathrm{obs}$, is less than half frequency bin $\delta f=1/2T_\mathrm{FFT}$. The frequency variation for spinning neutron stars can be expressed through the second order term of a Taylor expansion as
\begin{equation}
\Delta f = \ddot{f}\frac{T^2_\mathrm{obs}}{2}.
\end{equation}
Hence, the condition for neglecting the second order spin-down is
\begin{equation}
\ddot{f} \le \frac{1}{T^2_\mathrm{obs}\cdot T_\mathrm{FFT}}.
\label{eq:ddot}
\end{equation}
This can be translated in a maximum value of the second order spin-down term. Let us consider a power law to describe the relation among the signal frequency and its first time derivative:
\begin{equation}
    \dot{f} \propto f^n,
    \label{eq:brak}
\end{equation}
where $n$ is the \textit{braking index}, which value depends on the mechanism driving the rotational evolution of the star. By integrating the equation, we find the well-known relation for the time dependency of the frequency:
\begin{equation}
f(t)=f_0\left(1+\frac{t}{\tau}\right)^\frac{1}{1-n},
\label{eq:foft}
\end{equation}
where $\tau=(1-n)\frac{\dot{f}_0}{f_0}$ is the characteristic spin-down age which, for values of the frequency and spin-down typical for spinning neutron stars, is much bigger than any reasonable observation time of gravitational wave detectors.
By deriving Eq. \ref{eq:foft} two times, we obtain the following expression for the second time derivative:
\begin{equation}
\ddot{f}=\frac{f_0}{\tau^2}\frac{n}{(1-n)^2}\left(1+\frac{t}{\tau}\right)^\frac{2n-1}{1-n}
\end{equation}
Neglecting the very weak time dependency of $\ddot{f}$ and assuming n=5, which holds for objects which spin-down is dominated by the emission of gravitational waves, we have the well known relation
\begin{equation}
\ddot{f}_0=5\frac{\dot{f}^2_0}{f_0}
\label{eq:f2dot}
\end{equation}
For each pair $(f_0,~\dot{f}_0)$ of the searched parameter space, we can then determine if the corresponding value of $\ddot{f}_0$ satisfies Eq. \ref{eq:ddot}. 
\begin{figure}
    \centering
    \includegraphics[width=\columnwidth]{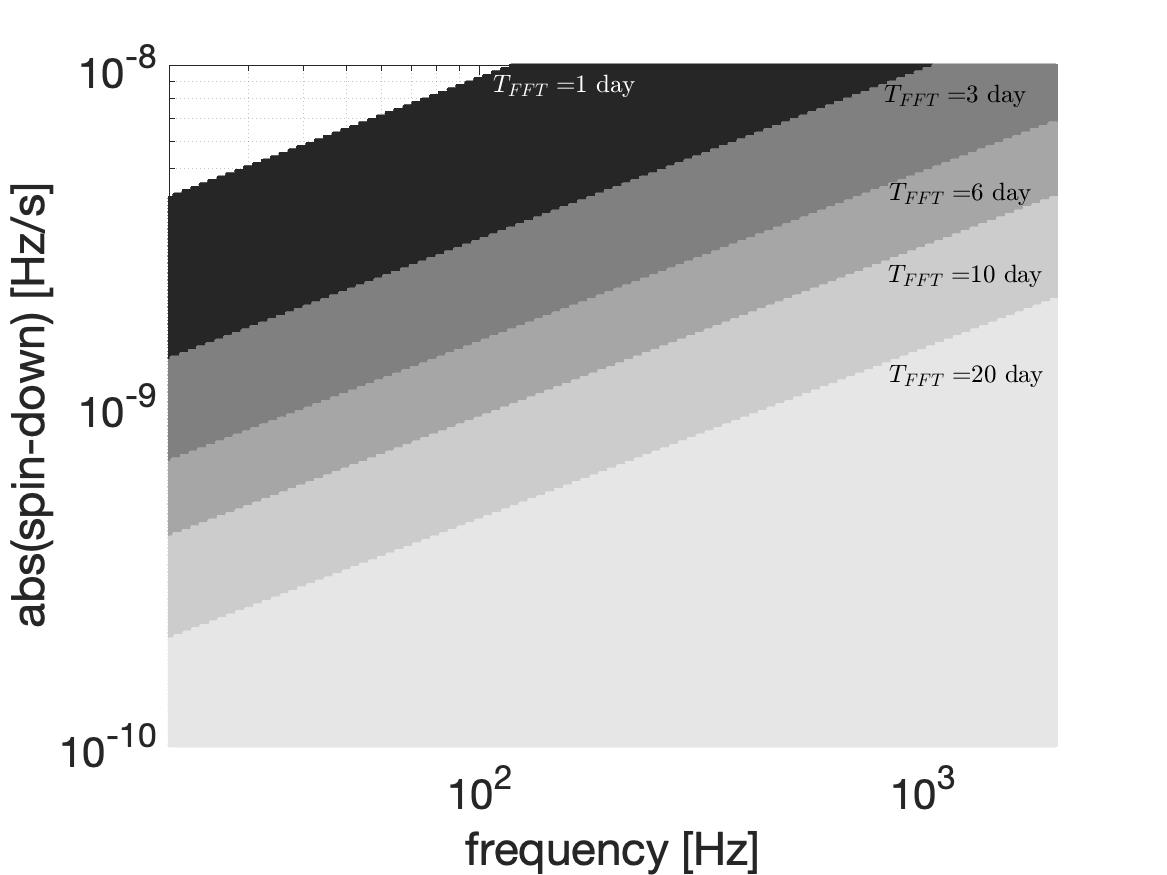}  
    \caption{For various values of $T_\mathrm{FFT}$, and $T_\mathrm{obs}=$ 1 yr, the plot shows the portion of parameter space for which the second order spin-down can be neglected.}
    \label{fig:f2dot}
\end{figure}
Fig. \ref{fig:f2dot} shows, for various values of $T_\mathrm{FFT}$, and $T_\mathrm{obs}=$ 1 yr, the portion of parameter space, defined by $f_0\in [20,~2000]$ Hz and $|\ddot{f}_0|\le 10^{-8}$ Hz/s, for which the second order spin-down can be neglected. The upper left white corner of the plot, corresponding to small signal frequency and very high spin-down, is the region for which the second-order spin-down is never negligible as soon as $T_\mathrm{FFT}\ge$ 1 day. On the other hand, as an axample a source emitting a signal at $f_0=$ 100 Hz, and analyzed dividing the data in segments of duration $T_\mathrm{FFT}=$ 6 days, could be searched neglecting the second order spin-down only if its first order spin-down was $|\dot{f}_0|<1.55\times 10^{-9}$ Hz/s. For each value of $T_\mathrm{FFT}$, the parameter space cut (the inclined straight line separating regions of different color) corresponds to a specific value of the second order spin-down, according to Tab. \ref{tab:f2dotvalues}, which is the maximum allowed value not requiring an explicit correction.
\begin{table*}[t]
    \begin{tabular}{|c|c|}
    \toprule
    $T_\mathrm{FFT}$ [days] & $\ddot{f}_{0} ~ [Hz/s^2]$ \\ \hline
  1 &  $4.25\times 10^{-18}$ \\
  3 & $4.72\times 10^{-19}$ \\
  5 & $1.18\times 10^{-19}$ \\
  10 & $4.25\times 10^{-20}$ \\
  20 & $1.06\times 10^{-20}$ \\
    \hline
    \end{tabular}
  \caption{Maximum allowed value of the second order spin-down which does not require an explicit correction, as a function of $T_\mathrm{FFT}$ and assuming $T_\mathrm{obs}=$ 1 yr. Such value corresponds to the separation line among different colored regions in Fig. \ref{fig:f2dot}.} 
    \label{tab:f2dotvalues}
\end{table*}

Two comments are in order. First, assuming a different spin-down mechanism, that is a different value for the \textit{braking index} in Eq. \ref{eq:brak}, would affect the position of the cuts in Fig. \ref{fig:f2dot} and the corresponding maximum allowed second order spin-down values of Tab. \ref{tab:f2dotvalues}. In general, when the spin-down of a spinning neutron star has non-GW contributions, the resulting \textit{braking index} is smaller than 5 (e.g. it is equal to 3 for pure dipole EM emission). This results in a smaller $\ddot{f}$ for given values of $(f,\dot{f})$. As a consequence, the allowed regions shown in Fig. \ref{fig:f2dot} are conservative. 

Second, known pulsars typically have second order spin-down values smaller than the values shown in the Table. Specifically, there is only one known pulsar, J0534+2200, with $\ddot{f}\simeq 1.11\times 10^{-20}~ [Hz/s^2]$, for which the correction for the second order spin-down would be needed if $T_\mathrm{FFT}\ge 20$ days. While properties of the unknown neutron star population could not be directly related to that of  known pulsars, we may expect that much larger second order spin-downs should not be extremely common. 

Nevertheless, the extension of the analysis method to include the correction of the second order spin-down will be an important step that we defer to future work.

\bibliographystyle{unsrt} 
\bibliography{references}
\end{document}